\documentclass[preprint,3p,12pt]{elsarticle}

\usepackage{amssymb}
\usepackage{amsmath}
\usepackage{hyperref}
\hypersetup{hidelinks}

\usepackage{booktabs}
\usepackage{tabularx}

\usepackage{supertabular}
\usepackage{amssymb}
\usepackage[super]{nth}
\journal{Physica A}
\usepackage[table,xcdraw]{xcolor} 
\usepackage{array}
\usepackage{cleveref}
\usepackage{float}

\usepackage[caption=false,font=footnotesize]{subfig}

\usepackage{color}
\usepackage{rotating}
\usepackage{tcolorbox}
\usepackage{float}
\tcbuselibrary{skins}

\usepackage{mathtools}

\DeclarePairedDelimiter\cardinality{\lvert}{\rvert}
\newcommand{\card}{\cardinality}

\DeclarePairedDelimiter\beliefbracket{\{}{\}}
\DeclarePairedDelimiter\statebracket{\langle}{\rangle}

\newcommand{\fusionrate}{\ensuremath{\sigma}}

\newcommand{\evidencerate}{\ensuremath{\rho}}

\newcommand{\errorrate}{\ensuremath{\epsilon}}
\newcommand{\error}{\errorrate}

\newcommand{\languagesize}{\ensuremath{n}}

\newcommand{\fusionThreshold}{\ensuremath{\gamma}}
\newcommand{\fthreshold}{\fusionThreshold}

\newcommand{\belief}{\ensuremath{B}}%beliefs
\newcommand{\beliefvector}{\ensuremath{\mathbf{P}}}%belief
\newcommand{\fmatrix}{\ensuremath{\mathbf{F}}}%Fusion matrix
\newcommand{\umatrix}{\ensuremath{\mathbf{U}}}%Updating Matrix

\newcommand{\f}{\ensuremath{\fmatrix}}%Fusion matrix
\newcommand{\updating}{\ensuremath{\umatrix}}%Updating Matrix

\newcommand{\state}{\ensuremath{s}}%states

\newcommand{\allstates}{\ensuremath{\mathbb{S}}}

\newcommand{\truestate}{\ensuremath{\state^*}}

\newcommand{\jaccard}{\ensuremath{J}}
\newcommand{\accuracy}{\ensuremath{\alpha}}

\newcommand{\intersect}{\ensuremath{\cap}}
\newcommand{\union}{\ensuremath{\cup}}
\newcommand{\given}{\ensuremath{\vert}}

\renewcommand{\leq}{\leqslant}
\renewcommand{\geq}{\geqslant}

\usepackage{xparse} % Required for optional arguments

\NewDocumentCommand{\figureAddLabel}{m m m m O{0.05} O{0.04}}{%
    \begin{tikzpicture}
        \node[anchor=south west, inner sep=0] (image) at (0,0) {\includegraphics[width=#1\linewidth]{#2}};
        \begin{scope}[x={(image.south east)},y={(image.north west)}]
            \IfNoValueF{#3}{\node[anchor=north] at (0.5, #5) {\scriptsize{#3}};} % Default 0.05
            \IfNoValueF{#4}{\node[anchor=east] at (#6, 0.5) {\scriptsize{#4}};} % Default 0.04
        \end{scope}
    \end{tikzpicture}%
}
\definecolor{myblue}{RGB}{71, 94, 209}
\newcommand{\revision}{\textcolor{black}}

\definecolor{royalblue}{rgb}{0.0, 0.14, 0.4}
\newcommand{\marginLeft}[1]{\text{\rotatebox{90}{\sffamily\bfseries\color{royalblue}\large{\phantom{j}#1}}}}
\newcommand{\bottomText}[1]{\text{{\sffamily\bfseries\color{royalblue}\large{\phantom{j}#1}}}}
\usepackage{booktabs}

\begin{document}

\begin{frontmatter}



\title{Imprecise Belief Fusion Improves Multi-agent Social Learning}

\author[label1]{Zixuan Liu}
\author[label1]{Jonathan Lawry}
\affiliation[label1]{organization={School
of Engineering Mathematics and Technology, University of Bristol},
             city={Bristol},
             postcode={BS8 1TW},
             country={UK}}
\author[label2]{Michael Crosscombe}
\affiliation[label2]{organization={Graduate School of Arts and Sciences, The University of Tokyo},
             city={Tokyo},
             postcode={153-8902},
             country={Japan}}             

% \author{Zixuan Liu, Jonathan Lawry, Michael Crosscombe} %% Author name

% \affiliation{organization={},%Department and Organization

%% Abstract
\begin{abstract}
In social learning, agents learn not only from direct evidence but also through interactions with their peers. We investigate the role of imprecision in such interactions and ask whether it can improve the effectiveness of the collective learning process. To that end we propose a model of social learning where beliefs are equivalent to formulas in a propositional language, and where agents learn from each other by combining their beliefs according to a fusion operator. The latter is parametrised  so as to allow for different levels of imprecision, where a more imprecise fusion operator tends to generates a more imprecise fused belief when the two combined beliefs differ. In this context we describe both difference equation models and agent-based simulations of social learning under a variety of conditions and with different initial biases. The results presented suggest that for populations with a strong initial bias towards incorrect beliefs some level of imprecision in fusion can improve learning accuracy across a range of learning conditions. Furthermore, such benefits of imprecision are consistent with a stability analysis of the fixed points of the proposed difference equation models.
\end{abstract}


\begin{highlights}
\item Study introduces a novel model of social learning where beliefs are represented as propositional formulas and combined through parameterised fusion operators that control belief imprecision.
\item Mathematical analysis and simulations reveal that when populations have strong initial biases toward incorrect beliefs, incorporating some imprecision in belief fusion can improve collective learning outcomes.
\end{highlights}

%% Keywords

\end{frontmatter}

% \linenumbers
\section{Introduction}
Social learning has been the subject of extensive research across various disciplines including biology, psychology, and more recently, multi-agent artificial intelligence and swarm robotic systems. It is common in social animals where individuals learn collectively by both observation and imitation of others~\citep{heyes1994social}, and it also plays an important role in human societies~\cite{boyd2011cultural}. 
\revision{Social learning allows individuals to acquire locally adaptive behaviours at a lower cost than individual learning. By learning through observation or imitation, individuals avoid the trial-and-error process required to independently innovate separate solutions to complex problems. This reduces the cost of learning across the population, as highlighted in~\citep{barrett2007hominid} which suggests that cognitive mechanisms underlying cultural transmission distribute the cost of acquiring adaptive information across many individuals and generations.
Social learning therefore  facilitates cumulative cultural evolution (CCE) by allowing individuals to acquire and refine knowledge from others, bypassing costly individual trial-and-error and cumulating small, successive improvements to be passed across generations.}  %Humans acquire cultural traditions from family members and students not only learn from teachers but also engage in group projects and discussions with classmates, thus acquiring collaborative skills and diverse perspectives. 
\revision{For example, the Inuit’s development of kayak keels demonstrates how innovations can arise through trial and error (e.g., experimenting with rudders) and subsequently spread through imitation, allowing gradual improvements over 50 years~\citep{boyd2011cultural}. CCE is a cornerstone of human adaptability, enabling the accumulation of innovations across generations through social learning. Boyd and Richerson introduced several models showing that social learning is adaptive in stable environments but requires balancing with individual learning under variable conditions~\citep{boyd2013evolutionary}. They emphasize that reliance on imitation alone risks maladaptive outcomes if models are outdated, highlighting the need for an optimal mix of learning strategies. Henrich and McElreath expanded on this approach by identifying cognitive mechanisms such as prestige bias (learning from successful individuals) and conformity bias (adopting majority behaviours), which enhance the fidelity and efficiency of cultural transmission~\citep{henrich2003evolution}. These biases reduce cognitive costs and enable populations to aggregate information effectively. }

Social insects, such as ants, bees, and termites, use social learning to enhance their collective efforts. They excel at coordinating large groups, whether constructing complex habitats or finding the most efficient paths for gathering resources. For instance, a single ant might seem very limited in its capabilities, but a colony, through the power of social learning and self-organisation, can construct architectural marvels, wage wars, or optimise foraging for food in surprisingly efficient ways. Ants, for instance, employ pheromone trails as a form of environmental marker to communicate with their peers about food sources or danger zones. This chemical communication allows them to collectively make decisions about the shortest paths to food or to mobilise against threats~\citep{holldobler1990ants}. Likewise, honeybees use the ``waggle dance" as a form of information sharing by which foraging bees can relay information regarding the location, distance, and quality of food sources to their hive mates, ensuring the efficient allocation of foragers to plentiful nectar sources~\citep{frisch1993dance}. Such achievements highlight the strength of social learning, as a means of optimising collective behaviours.  For example, honeybee foragers are guided by social cues related to floral scents, conveyed through trophallaxis. This process involves successful foragers transferring nectar samples to nestmates, thereby disseminating information about nectar sources within the hive \citep{frisch1993dance}.  In addition, one of the most significant challenges encountered by social insect colonies is relocating their nests. This complex process involves the coordinated movement of hundreds or thousands of individuals from their current nest to the best new site as selected from several possible alternatives. The decision-making process is driven by a few scout members who, through a mix of individual and social learning, manage to achieve consensus on the optimal site \citep{franks2003speed}. Remarkably, this consensus is reached despite most individuals not evaluating all possible locations themselves. Artificial social learning algorithms, inspired by natural systems, have been implemented in synthetic autonomous systems, such as robotic swarms, enabling consensus formation in various contexts, including the selection of an aggregation site~\citep{sion2022controlling}, choosing a common direction of movement~\citep{onur2024predictive}, or identifying the dominant environmental features ~\citep{shan2021discrete}.

Agents in cooperative systems learn both from direct evidence in the environment and from the beliefs of fellow agents through a process of belief fusion. There are several well-established fusion strategies applied to social learning and collective decision-making problems. A common type of strategies are voting-based methods. The majority rule, for instance, is a straightforward method where an agent adopts the most commonly held belief amongst its neighbours as a new belief \cite{valentini2016collective}. Although simple, this method can be effective under certain conditions, especially when the population of agents is large and individual beliefs are reasonably accurate. Another popular voting-based approach is the weighted voter model \citep{valentini2014self,crosscombe2017robust}. In contrast to the majority rule, the weighted voter model is probabilistic, assigning varying degrees of probability to different beliefs based on the number of votes for each, i.e. the proportion of connected agents that holds that belief. In this context truth value-based fusion models have been shown to be more robust to noise than the weighted voter model~\cite{crosscombe2017robust}. In addition to voting-based frameworks, agents may also engage in belief fusion with their peers by means of fusion operators, typically involving a fixed number of agents for fusion. For instance, in \citep{lee2021negative}, where beliefs are probabilities, agents update their beliefs based on Bayes' theorem and engage in belief fusion using the multi-hypothesis product operator \citep{dietrich2016probabilistic}. This approach is particularly suited to noisy environments in which agents often receive erroneous quality values for different options. Similarly, in belief models based on Possibility Theory or Dempster-Shafer Theory, pairwise fusion operators can be employed~\citep{crosscombe2019distributed,crosscombe2019evidence}. Possibility theory can be interpreted as an imprecise version of probability theory incorporating upper and lower probabilities, and has been shown to outperform a similar probabilistic model for social learning when evidence is scarce  \citep{crosscombe2019distributed}. For Dempster-Shafer theory the robustness to evidential noise of several fusion operators has been compared when applied to the best-of-$n$ problem~\citep{crosscombe2019evidence}. In particular, the best performance under noisy conditions was achieved by Yager's operator~\citep{yager1992specificity} and by Dubois \& Prade's operator~\citep{dubois1988representation} rather than by the more commonly applied Dempster's fusion rule. \revision{Furthermore, ~\citep{Bartashevich21} investigated the performance of a range of different Dempster-Shafer theory fusion operators applied to multi-feature collective perception problems. In this context results suggest that fusion which involves some form of proportional conflict redistribution leads to good collective performance.}

The exploration of imprecision in social learning is to some extent inspired by collective behaviour in nature where populations inhabit hostile and rapidly changing environments about which there is only limited and noisy information. In such scenarios biological systems can often exploit errors, imprecision and disturbances to enhance their collective performance. In particular, empirical studies of social insects reveal that these organisms exploit imprecision to their advantage during tasks such as food source selection. This phenomenon has been observed in both honeybees and ants. For example, the honey bee waggle dance exploits imprecision in the way it indicates directions, so that recruits are more spacially spread-out as is consistent with the ``tuned error" hypothesis~\citep{towne1988spatial, okada2014error}. Ants also utilise variability in trail-following behaviour, allowing them to respond effectively to environmental changes~\citep{dussutour2009noise}. This suggests that populations can sometimes collectively harness inherent imprecision and noise to positive effect~\citep{meyer2017role}. The property of benefiting from disturbances is sometimes referred to as antifragility~\citep{axenie2024antifragility} and is important in the design of artificial collective systems which, rather than being limited by the inevitable noise and imprecision present in a system, can instead attempt to exploit it so as to enhance robustness and performance. For example, recent studies of antifragility in collective systems have shown that systems with fusion noise can provide better performance than noise-free systems~\citep{zakir2024miscommunication}. Similarly, other work has shown that limited communication~\citep{mohamed2021when} or slower dynamics~\citep{gershenson2015slower} can improve the system's ability to spread new information and to react quickly to changes. In our previous work, fusion imprecision has also been shown to have the potential to improve learning accuracy in the presence of sensor noise~\citep{liu2021imprecise}. Despite recent advances in the study of antifragility, the majority of research remains focused on agent-based simulations or robotic experiments, leaving analytical questions unanswered. 

%In this context, it is therefore interesting to investigate if artificial multi-agent systems can adapt to imprecision and whether some forms of imprecise belief fusion may even enhance learning in the presence of sensor error or other types of environmental noise. 

Throughout the literature on social learning and opinion dynamics agents' beliefs are typically in the form of a single real number ~\citep{hegselmann2006truth} or a truth-value ~\citep{crosscombe2017robust}. However, in practice social learning problems are often complex requiring agents to take account of multiple features or propositions. This in turn necessitates a richer language in which to represent beliefs. Following ~\citep{riegler2009}, ~\citep{wenmackers2012} and  ~\citep{cholvy2018opinion} in this paper we assume that beliefs are equivalent to formulas in propositional logic which describe individuals current opinions about what is the true state of the world.

In a propositional logic setting we will propose a model in which the level of fusion imprecision can be controlled by a Jaccard similarity based parameter and then use this model to investigate whether artificial agents can benefit from imprecise fusion during social learning. More specifically, \revision{we} will introduce a difference equation-based model to investigate social learning behaviour under different levels of fusion imprecision. Agents' beliefs will be represented by sets of possible states, each corresponding to an allocation of truth values to the propositions in the language, and where the cardinality of a belief set is a measure of its imprecision. We apply the parameterised fusion operator introduced in~\citep{liu2021imprecise} and which returns beliefs of varying levels of imprecision. Imprecise fusion operators have been shown to be optimal when the frequency of fusion is high relative to the frequency with which noisy evidence is obtained from the environment. In this research we systemically investigated the fusion operator over a broader range of parameters with both agent-based and difference equation-based methods. Using this  approach, we investigate the effect on learning performance of varying the frequency of agent communication, the rate at which evidence is received, the degree of evidential noise, and the level of fusion imprecision.

The main contributions of the paper are as follows: We introduce a non-linear model for social learning with imprecise belief fusion operators that can handle varying levels of uncertainty and noise in collective decision-making. The study provides new theoretical insights into the role of imprecision in belief fusion and how it can improve robustness in the presence of noisy evidence. We show that when communications between agents are relatively infrequent this model of social learning is robust to varying initial conditions, noisy evidence and levels of fusion imprecision. Furthermore, we demonstrate that the proposed imprecise fusion model outperforms traditional precise models, especially for certain rates of evidence collection and agent interaction. 

The remainder of this paper is structured as follows:  \Cref{sec:social-learning-model}  introduces our proposed propositional model including the imprecise fusion operator and explains its relevance to collective decision-making. \Cref{sec:methods} defines a difference-equation model incorporating imprecise fusion and then \Cref{sec:fixed-point-analysis} presents an analysis of its fixed points under various learning scenarios and initial conditions. \Cref{sec:difference-equation-simulations} studies the different-equation model using simulation. In \Cref{sec:agent-based-simulation}, we describe the agent-based simulations used to evaluate the model’s performance under a variety of conditions and different population sizes. The results of these simulations along with a comparative analysis between precise and imprecise fusion approaches are then presented. \Cref{sec:conclusion} concludes the paper by summarising our key findings and discussing the broader implications of this work for future research and practical applications.

% \section{Model Formulation}

% •	Mathematical Background: Provide the necessary mathematical foundations, including the theory behind non-linear dynamics and its application in MAS. This might include concepts from differential equations, chaos theory, or bifurcation theory.
% •	Proposed Model: Present the new or adapted non-linear model. Clearly describe its structure, assumptions, and how it addresses the complexities of collective decision-making in MAS.
% •	Equations and Properties: Provide the key mathematical equations that define the model. Discuss important properties of the model, such as stability, convergence, and how it handles the non-linearities in the system.
\section{Social Learning Model}
\label{sec:social-learning-model}

In this section, we introduce a social learning model that can incorporate varying levels of imprecision in belief fusion. We first outline the set-based belief model in \Cref{sec:belief-represention}, followed by the fusion operator in~\Cref{sec:belief-fusion-operators}. We then introduce an evidential updating strategy in~\Cref{sec:evidential-udpating-method}

\begin{figure*}[h]
    \centering
    \includegraphics[width=0.9\linewidth]{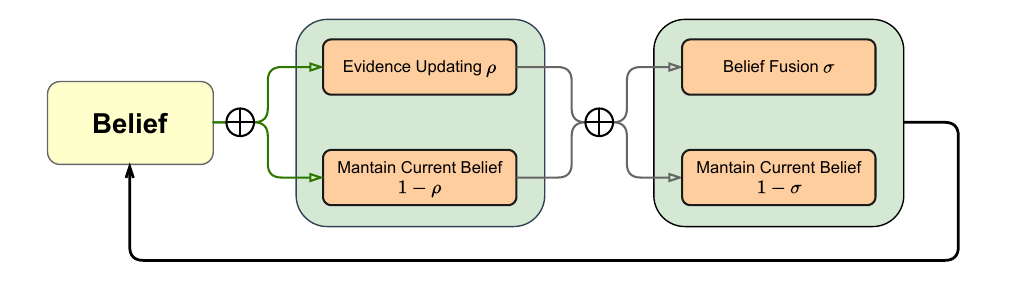}
    \caption{Diagram of an updating-fusion iteration. During the iteration, agents uncertain about propositions seek evidence with success rate $\evidencerate$ before all agents then attempt to fuse their beliefs with peers at a rate $\fusionrate$.}
    \label{fig:iteration-diagram}
\end{figure*}

\subsection{Belief Representation}\label{sec:belief-represention}

Consider a population of agents attempting to collectively learn the state of their environment, which we assume can be described by a finite set of propositions $\mathcal{P} = \{p_1,...,p_n\}$. We assume that the propositional language is fixed during the learning process. From this perspective a state $\state$ is the allocation of Boolean truth values to each of the propositions. In other words, a state is a function $\state : \mathcal{P} \rightarrow \{0, 1\}^n$. For notational convenience we represent a state $\state$ by the $\languagesize$-tuple $\statebracket{\state(p_1), \ldots, \state(p_n)}$. In this case if we know which of the propositions are true and which are false, then we know exactly what the state of the world is, at least as far as is expressible in the propositional language, and hence we refer to such truth-value allocations as possible worlds. 

Let $\allstates$ denote the set of all worlds, so that a language with $n$ propositions allows us to distinguish between $\card\allstates$ =
$2^n$ possible worlds. In particular, there is exactly one possible world in
which all propositions are true; there are $n$ possible
worlds in which all but one of the propositions are
true; there are $\binom{n}{2}$ possible worlds in which all but two
of the atomic sentences are true; and so on. For example, when there are $n = 2$ propositions, then there are
$\card\allstates = 4$ possible worlds $\mathbb{S}=\{s_1, \ldots, s_4\}$ as listed
in \Cref{tab:number-of-states}.
\newcolumntype{P}[1]{>{\centering\arraybackslash}p{#1}}
\begin{table}[t]
     \caption{With $n = 2$, there are $ 2^n = 4$ possible worlds, $\mathbb{S}=\{s_1, \ldots, s_4$\}.}
    \label{tab:number-of-states}
    \centering
    \begin{tabular}{P{0.2\linewidth} P{0.2\linewidth} P{0.2\linewidth}}
    \hline
        & $p_1$ & $p_2$ \\
        \hline
        $s_1$ & 1 & 1 \\
        $s_2$ & 1 & 0 \\
        $s_3$ & 0 & 1 \\
        $s_4$ & 0 & 0 \\
        \hline
    \end{tabular} 
\end{table}

An agent's belief $\belief \subseteq \allstates$ is then taken to be the set of states which the agent believes can possibly be the true state $\truestate$. Such beliefs can also represent opinions in the form of propositional logic formulas, where $B$ is the set of truth-value allocations for which the formula is true. $B$ can be any non-empty subset of $\allstates$, and hence there are $\card\allstates^n-1$ different possible beliefs. In this setting imprecise beliefs are those subsets of $\allstates$ with cardinality $\card\belief > 1$ while a singleton belief $\belief = \beliefbracket{\state}$ means that the agent is certain that $\state$ is the true state. We assume that agents adopt a closed-world assumption which in this context means assuming that $\allstates$ covers all possible states of the world. Therefore, agents' beliefs are constrained such that $B\neq\emptyset$ since it cannot be the case that all states in $\mathbb{S}$ are impossible. Note that a given belief $\emptyset \neq \belief \subseteq \allstates$ classifies each proposition $p_i$ as being either true, if $s(p_i)=1$ for all $\state \in \belief$, false, if $s(p_i)=0$ for all $\state$ in $\belief$, or uncertain otherwise. Hence, the more imprecise an agent's belief the more propositions they will tend to be uncertain about. This indicates a natural relationship between the set-based model of beliefs and three-valued approaches~\citep{crosscombe2017robust}. For example, consider a search and rescue scenario with $2$ locations. Then let $p_i$ denote the proposition `casualties are in location $i$' for $i=1,2$. Now consider the belief $\belief$ given by,
\begin{gather*}
\belief=\beliefbracket{\statebracket{1,0},\statebracket{0,1} }   
\end{gather*}
In this case $\belief$ corresponds to the belief that there are casualties either in location $1$ or location $2$ but not both, i.e. as a propositional logic formula this corresponds to the exclusive OR of $p_1$ and $p_2$. Note that, according to $\belief$, $p_1$ and $p_2$ are both uncertain.

\subsection{Belief Fusion Operators}\label{sec:belief-fusion-operators}

We now introduce a parameterised fusion operator which returns beliefs of varying levels of imprecision. This requires a measure of the similarity between beliefs for which we use the well-known Jaccard similarity \citep{jaccard1912distribution} defined as follows: For $\belief_1, \belief_2 \subseteq \allstates$,

\begin{equation}\label{eq:jaccard-sim}
  \jaccard (\belief_1,\belief_2) =\frac{\card{\belief_1\intersect\belief_2}}{\card{\belief_1\union\belief_2}}
  \end{equation}
  
We now define the similarity threshold fusion operator as follows: for $\fthreshold \in [0,1]$,
\begin{equation}\label{eq:gamma-fusion-operator}
{\belief_1}\odot_\fthreshold{\belief_2}=
	\begin{dcases}
        ~~{\belief_1}\intersect{\belief_2} &: ~~J({\belief}_1,{\belief}_2) > \fthreshold \\
        ~~{\belief_1}\union{\belief_2} &: ~~J({\belief}_1,{\belief}_2) \leq \fthreshold
	\end{dcases}
\end{equation}

%The operator still preserves the possibility and impossibility of both beliefs, i.e. still satisfies the property of unanimity. The rule is not optimistic because the resulting belief is not necessarily a subset of either belief when they overlap but not similar enough and therefore minimal commitment does not hold as well. However, the operator improves the rigidity and extremity of the original one and makes the transaction from intersection to union much smoother. 
For example, let $\belief_1=\beliefbracket{s_1,s_2,s_3}$ and $\belief_2=\beliefbracket{s_2,s_3,s_4}$ then,

\begin{gather*}
\jaccard(\belief_1,\belief_2)=  \frac{\card{\beliefbracket{s_2,s_3}}}{\card{\beliefbracket{s_1,s_2,s_3,s_4}}}=\frac{1}{2}  
\end{gather*}

and hence $\belief_1 \odot_\fthreshold \belief_2=\beliefbracket{s_2,s_3}$ if $\fthreshold < \frac{1}{2}$ and $\belief_1 \odot_\fthreshold \belief_2=\{s_1,s_2,s_3,s_4\}$ for 
$\fthreshold \geq \frac{1}{2}$.

Note that for $\fthreshold=0$ this operator corresponds to the intersection-union operator \cite{dubois1988representation} as given by:
\begin{gather}\label{eq:belief-fusion-intersection-union-operator}
    \belief_1 \odot_0 \belief_2 =
    \begin{dcases}
        ~~\belief_1 \intersect \belief_2 &: ~~\belief_1 \intersect \belief_2 \neq \emptyset \\
        ~~\belief_1 \union \belief_2 &: ~~\belief_1 \intersect \belief_2 = \emptyset
    \end{dcases}
\end{gather}
On the other hand, for $\fthreshold=1$ we have that $\belief_1 \odot_1 \belief_2 = \belief_1 \union \belief_2$. In general, $\fthreshold$ controls the level of generality or precision of the operator such that for $\fthreshold \leq \fthreshold^\prime$, $\belief_1 \odot_\fthreshold \belief_2 \subseteq \belief_1 \odot_{\fthreshold^\prime} \belief_2$ for all sets  $\belief_1, \belief_2 \subseteq \allstates$. 
%One of the limitations of the intersection-union operator is its discontinuity when the two beliefs transition from being consistent to inconsistent. For example, consider \( B_1 = \{s_1, \dots, s_i\} \), \( B_2 = \{s_i, \dots, s_n\} \), and \( B'_2 = \{s_{i+1}, \dots, s_n\} \). In this case, \( B_1 \odot_0 B_2 = \{s_i\} \), whereas \( B_1 \odot_0 B'_2 = \mathbb{S} \). In other words, the operator may sometimes produce an extreme belief that does not accurately reflect the initial beliefs being combined. The imprecise fusion operator introduced in \Cref{eq:gamma-fusion-operator} addresses this issue by incorporating a threshold $\gamma$ on the Jaccard similarity between the two beliefs. Specifically, the beliefs are only intersected if their degree of agreement is sufficiently high, i.e.\ $J(B_1, B_2) > \gamma $.
\revision{Precise fusion, i.e. the case where $\gamma=0$ in \Cref{eq:belief-fusion-intersection-union-operator}, is a form of logic-based merging ~\citep{konieczny2002merging} that has been extensively studied in a variety of contexts including as an alternative operator in Dempster-Shafer theory \citep{dubois1988representation}. It captures the intuition that if two beliefs are consistent then combining them should reinforce belief in those states about which there is agreement, while if they are inconsistent then fusion should increase uncertainty about what is the true belief bounded by the assumption that the latter is contained in at least one of the two belief sets being combined.
More formally, we can analyse the operator in terms of the fusion principles introduced in \citep{dubois2016basic}. Of particular relevance are the following three principles: \emph{Optimism} which requires that the combination of two intersecting belief sets is a subset of both of them; \emph{Unanimity} according to which the combination of $B_1$ and $B_2$ is a superset of $B_1 \cap B_2$ and a subset of $B_1 \cup B_2$; \emph{Minimal commitment} which requires that the combination of $B_1$ and $B_2$ is the most imprecise belief set satisfying optimism and unanimity. Clearly these three principles are sufficient to uniquely characterise the precise fusion operator. However, to justify the imprecise fusion operator, $\odot_\gamma$ requires a variation of the optimism principle which we might refer to as \emph{bounded optimism}; the combination of two sets $B_1$ and $B_2$ is a subset of them both provided that $B_1$ and $B_2$ are \emph{sufficiently similar} intersecting sets. In this case bounded optimism, unanimity and minimal commitment characterise $\odot_\gamma$ provided that we interpret the term \emph{sufficiently similar} to mean that $J(B_1,B_2) \geq \gamma$.}

The use of Jaccard similarity to deal with inconsistency has also been proposed by~\citep{schockaert2010inconsistency} who applies similarity-based enlargement of the sets of interpretations to resolve inconsistencies in fusion problems. Note that by applying a finite set, the number of possible similarities between beliefs is also finite. For example, in the case of $n=2$ propositions, there are $4$ possible states, resulting in $7$ potential similarity values corresponding to $\{0,~\frac{1}{4},~\frac{1}{3},~\frac{1}{2},~\frac{2}{3},~ \frac{3}{4},~1\}.$

\subsection{Evidential Updating Method}\label{sec:evidential-udpating-method}

In addition to fusing with other agents, agents receive direct information from the environment as evidence. For evidential updating, we assume that the evidence takes the form of an assertions about the true state of the world $\truestate$. In the current model such an assertion identifies a set of states $E$ that can possibly be the true state. 

Given $E$ we then propose that an agent updates their belief $\belief$ to $\belief\given E$ such that:

\begin{gather}
    \label{eq:evidence-update-intersect-only}
    B \vert E = 
    \begin{cases} 
    B \cap E:B \cap E \neq \emptyset
    \\ B:\text{otherwise.} 
    \end{cases} 
\end{gather}

This method of evidential updating in which certain states are ruled out as part of the learning process has already been applied effectively in social learning for best-of-$n$ problems~\citep{lawry2019epistemic}. In the sequel we will assume that evidential assertions relate to the propositions for the language, so that for some $p_i \in \mathcal{P}$, either $E=\{\state:\state (p_i)=1\}$ or $E=\{\state:\state (p_i)=0\}$.

\subsection{The Iterative Learning Process}\label{sec:iteration-model}

We investigate social learning behaviour as a population of agents attempts to reach a consensus about a propositional state description of the world through evidential updating and belief fusion, where both processes may be limited. \Cref{fig:iteration-diagram} is a flow diagram showing how an agent's belief is stochastically updated during a single iteration in which they first carry out evidential updating and then belief fusion. For evidential updating an agent receives evidence about the propositions in the language with probability $\rho$ (the \emph{evidence rate}) in which case it updates its belief according to~\Cref{eq:evidence-update-intersect-only}. However, with probability $1-\rho$ it receives no evidence and its belief remains unchanged at this stage. The agent will stop collecting evidence if it is certain about every proposition, i.e.\ if its belief set is a singleton. Following the evidential updating step an agent will engage in belief fusion with probability $\sigma$ (the \emph{fusion rate}). This means it will enter a pool of agents to be randomly paired for belief fusion using~\Cref{eq:gamma-fusion-operator}. For every pair, both agents will adopt the result of this fusion as their new belief. If the number of agents in the pool is odd, then one agent will not take part in fusion. Such pairwise belief fusion is applied in many similar models, such as in \cite{lawry2019epistemic,crosscombe2021impact}.

\revision{The parameter values for $\fusionrate$ (fusion rate) and $\evidencerate$ (evidence rate) can be selected to capture a broad spectrum of dynamics reflecting constraints on individual learning, i.e. evidential updating, and social learning, i.e. belief fusion. The evidence rate models the availability or accessibility of evidence in the environment so that low values of $\rho$ close to $0$ reflect a sparsity of evidence, i.e. as in isolated or resource-constrained environments, whereas high values close to $1$ reflect an abundance of evidence, i.e as in structured, data-rich environments. The fusion rate parameter $\sigma$ provides a basic model of communication constraints within collective systems, reflecting factors such as limited bandwidth, resource conservation, signal interference, network scalability, and security concerns, which are all prevalent in real-world scenarios. Low values of $\sigma$ close to $0$ model infrequent interactions or scenarios with limited social exchange, i.e. as for isolated individuals or where there is highly constrained communication, whereas high values of $\sigma$ close to $1$ model frequent and widespread social interactions, i.e. as in highly connected social networks with unconstrained communication.}

%The fusion rate parameter provides a basic model of communication constraints within collective systems, reflecting factors such as limited bandwidth, resource conservation, signal interference, network scalability, and security concerns, which are all prevalent in real-world scenarios. The fusion rate also serves as a control mechanism for the frequency of belief updates. Limiting the frequency of fusion, as identified in recent research \citep{crosscombe2023benefits}, can result in improved robustness to noise in evidence. As shown in \Cref{fig:iteration-diagram}, taking account of $\rho$ and $\sigma$ there are $4$ possible updating and fusion outcomes for an agent. For example, with probability $\evidencerate\times\fusionrate$ the agent will both receive evidence and enter the fusion pool.  

Another key parameter is error rate $\error$ which is introduced as a simple model of environmental or sensor errors. More specifically, we define the error rate as the probability $\error$ that the evidence received is incorrect. More formally, when an agent receives evidence then we assume that they independently obtain information about the truth value of each proposition, which is correct with probability $1-\epsilon$ and incorrect with $\epsilon$. In other words, the evidence received has the form,
 \begin{gather*}
E=\bigcap_{i=1}^n E_i \text{ where } E_i=\{s:s(p_i)=s_i\}
\end{gather*}
and where $s_i=s^*(p_i)$ with probability $1-\epsilon$ and $s_i=1-s^*(p_i)$ with probability $\epsilon$. Hence, for $n=2$ the possible evidence assertions an agent can receive are $E=\{s^*\}$, $E=\{s_2\}$, $E=\{s_3\}$ and $E=\{s_4\}$, with respective probabilities $(1-\epsilon)^2$, $\epsilon(1-\epsilon)$, $(1-\epsilon)\epsilon$ and $\epsilon^2$. 

\section{Methodology}\label{sec:methods}
Based on the models we proposed above, we use two methods to study social learning behaviour; a difference equation based analytical model and agent-based simulations.

\subsection{Difference Equation Models}
In the following we propose a model for social learning behaviours. We focus on a language with $2$ propositions, where agents may hold $15$ different possible beliefs. For the iterative model proposed in \Cref{sec:iteration-model}, difference equation models can then be formulated in terms of the proportion of agents in a large population holding each belief. More formally, the social learning model can be represented as follows: Let

\begin{gather}
    \beliefvector_t = [P_t(1) ,P_t(2), \dots ,P_t(15)]
\end{gather}

denote the proportions of the each belief state in the agent population at iteration $t$. Without loss of generality we assume that $P_t(1)$ is the proportion of agents with a precise belief about the true state, i.e $B=\{\truestate\}$, and $P_t(15)$ is the proportion of completely ignorant agents with belief $B=\mathbb{S}$. The belief states corresponding to each index are listed in \Cref{tab:belief-notations}.

% \begin{gather*}
% \truestate=\statebracket{1,1}, s_2=\statebracket{1,0}, s_3=\statebracket{0,1}, 
% s_4=\statebracket{0,0};
% \\
% P_t(1)=\beliefbracket{\truestate},~
% P_t(2)=\beliefbracket{s_2},~
% P_t(3)=\beliefbracket{s_3},~
% P_t(4)=\beliefbracket{s_4},~
% \\
% P_t(5)=\beliefbracket{\truestate,s_2},~
% P_t(6)=\beliefbracket{\truestate,s_3},~
% P_t(7)=\beliefbracket{\truestate,s_4},~
% \\
% P_t(8)=\beliefbracket{s_2,s_3},~
% P_t(9)=\beliefbracket{s_2,s_4},~
% P_t(10)=\beliefbracket{s_3,s_4},~
% \\
% P_t(11)=\beliefbracket{\truestate,s_2,s_3},~
% P_t(12)=\beliefbracket{\truestate,s_2,s_4},~
% \\
% P_t(13)=\beliefbracket{\truestate,s_3,s_4},~
% P_t(14)=\beliefbracket{s_2,s_3,s_4},~
% \\
% P_t(15)=\beliefbracket{\truestate,s_2,s_3,s_4},~
% \end{gather*}
\begin{table}[ht]
    \centering
    \caption{Table of indices for the different belief state}\label{tab:belief-notations}
    \begin{tabular}{|c|c|c|c|}
\hline
index $i$ & $\belief_i$ & index $i$ & $\belief_i$ \\
\hline
1 & $\{s^*\}$ & 9 & $\{s_2, s_4\}$ \\
2 & $\{s_2\}$ & 10 & $\{s_3, s_4\}$ \\
3 & $\{s_3\}$ & 11 & $\{s^*, s_2, s_3\}$ \\
4 & $\{s_4\}$ & 12 & $\{s^*, s_2, s_4\}$ \\
5 & $\{s^*, s_2\}$ & 13 & $\{s^*, s_3, s_4\}$ \\
6 & $\{s^*, s_3\}$ & 14 & $\{s_2, s_3, s_4\}$ \\
7 & $\{s^*, s_4\}$ & 15 & $\{s^*, s_2, s_3, s_4\}$ \\
8 & $\{s_2, s_3\}$ &  &  \\
\hline 
\multicolumn{4}{|c|}{$\truestate=\statebracket{1,1}, s_2=\statebracket{1,0}, s_3=\statebracket{0,1}, s_4=\statebracket{0,0}$}\\
\hline
\end{tabular}
\end{table}

At a macro level, social learning can then be modelled as a difference equation of the form:
\begin{gather}
    \beliefvector_{t+1}^\intercal=\updating\f^{\beliefvector_t}_\fthreshold \beliefvector_t^\intercal
\end{gather}

where $\updating$ and $\f^{\beliefvector}_\fthreshold$ are matrices of transition probabilities for
the updating and fusion processes respectively. These are $15\times15$ matrices in the form: 
\begin{gather}
% \begin{pmatrix}
% P(i|j)
% \end{pmatrix}_{i,j=1}^{15}
\begin{pmatrix}
P(1|1) & P(1|2) & P(1|3) & \dots & P(1|15) \\
P(2|1) & P(2|2) & P(2|3) & \dots & P(2|15) \\
P(3|1) & P(3|2) & P(3|3) & \dots & P(3|15) \\
\vdots & \vdots & \vdots & \ddots & \vdots \\
P(15|1) & P(15|2) & P(15|3) & \dots & P(15|15)
\end{pmatrix}
\end{gather}

Here $P(j|i)$ represents the probability that an agent currently holding belief state $i$ will transition to belief state $j$. In general, the transition probabilities for the fusion process depend on both the fusion threshold $\fthreshold$ and the current distribution of belief states within the population, as they take account of the likelihood of an agent interacting with another agent who holds any of the various belief states, i.e. the fusion process is non-linear. This is indicated by the superscript in $\f^\beliefvector_\gamma$. For example, suppose that the fusion imprecision parameter is $\gamma=\frac{1}{3}$ then consider the fusion transition probability $P(13|6)$ quantifying the probability that an agent currently with belief $\{s^*,s_3\}$ will change their belief to $\{s^*,s_3,s_4\}$ as a result of the fusion process. Now in this case such a transition can only occur if the agent pools their belief with another agent currently holding one of the following beliefs; $\{s_4\}$,  $\{s^*,s_4\}$ or $\{s_3,s_4\}$. This has probability $\sigma \left( P(4) + P(7)+ P(10) \right)$. More generally, for $\f^{\beliefvector_t}_\fthreshold$ we have that:

\begin{gather}
P(j|i)=\sum_{B_k:B_k \odot_\gamma B_i=B_j} \sigma P(k) + \delta_{i,j} (1-\sigma)P(i)
\end{gather}

where \( \delta_{ij} \) is the Kronecker delta, which equals 1 when \( i = j \), and 0 otherwise. The full fusion transition matrix is given in \revision{\ref{ap:transition-matrices}} for $\fthreshold=0$.

In contrast to fusion, the transition probabilities for the evidential updating process ($\updating$) do not depend on the current proportions of beliefs. Furthermore, the evidential transition matrix $\updating$ is sparse because there are only a limited number of possible transitions. For example, for agents with current beliefs $\belief_1$ to $\belief_4$ evidence updating will not result in a change to their beliefs as they are certain about the state of the world. This means that for $i\leq4$, $P(j|i) = 1~if~i= j \text{ and } P(i|j) = 0$ otherwise. For beliefs $B_i$ where $i\in[5,10]$ so that they contain $2$ states, updating can result in a transition to one of three $3$ possible $B_j$, including $B_i$ itself and its subsets. For example, suppose that an agent has belief $B_5=\{s^*,s_2\}$ then evidential updating there are five possibilities; 1) with probability $1-\rho$ the agent does not receive any evidence and its belief remains as $B_5=\{s^*,s_2\}$; 2) with probability $\rho(1-\epsilon)^2$ it receives evidence $E=\{s^*\}$ and hence updates its belief to $B_1=\{s^*\}$; 3) with probability $\rho(1-\epsilon)\epsilon$ it receives evidence $E=\{s_2\}$ and hence updates its belief to $B_2=\{s_2\}$; 4) with probability $\rho\epsilon(1-\epsilon )$ it receives evidence $E=\{s_3\}$ and hence leaves it belief unchanged $B_5=\{s^*,s_2\}$; 5) with probability $\rho\epsilon^2$ it receives evidence $E=\{s_4\}$ and hence leaves it belief unchanged $B_5=\{s^*,s_2\}$. Taking account of these possible outcomes together with their associated probabilities we obtain the following transition probabilities: 
\begin{gather*}
P(5|5)=1-\rho + \rho \epsilon (1-\epsilon)  +\rho \epsilon^2=1-\rho(1-\epsilon)\\
P(1|5)=\rho(1-\epsilon)^2 \\
P(2|5)=\rho(1-\epsilon)\epsilon\\
P(j|5)=0 \text{ for } j \neq 1,2,5
\end{gather*}
The full transition matrix for updating can be found in  \ref{ap:transition-matrices}. 

\section{Fixed Point Analysis for Varying Levels of Fusion Imprecision}\label{sec:fixed-point-analysis}

In this section, we investigate the effect of different levels of fusion imprecision by identifying the equilibrium (fixed) points of the relevant difference equations and analysing their stability. For notational convenience we let $\beliefvector^t=(1,0,0,0,\mathbf{0})$, $\beliefvector^{h1}=(0,1,0,0,\mathbf{0})$, $\beliefvector^{h2} = (0,0,1,0,\mathbf{0})$, and $\beliefvector^f=(0,0,0,1,\mathbf{0})$ where $\mathbf{0}$ denotes a string of $11$ zeros. These are all fixed points since they correspond to scenarios in which the population have reached consensus on a precise belief, i.e. $\{s^*\}$, $\{s_2\}$, $\{s_3\}$ and $\{s_4\}$ respectively, and hence all agents will stop updating from evidence because their beliefs are precise, and fusion will have no effect since they all hold the same belief.  Furthermore, these are the only fixed points since any difference in beliefs between agents will result in further fusion, and any belief imprecision will result in further evidential updating. $\beliefvector^{h1}$ and $\beliefvector^{h2}$ correspond to the situation in which an only partially correct consensus has been reached. Specifically, in the case of $\beliefvector^{h1}$ the consensus is that proposition $p_1$ is true and $p_2$ is false, while in the case of $\beliefvector^{h2}$ $p_1$ is false and $p_2$ is true. By symmetry of the model these two fixed point have identical properties and hence we will not distinguish between them, instead using the notation $\beliefvector^{h}$ to stand for either. The stability of these three fixed points can now be determined from the Jacobian matrix for the difference equation in the normal manner. Specifically, the eigenvalues are determined for the Jacobian at a given fixed point and if these values are all strictly less than $1$, then the equilibrium point is stable, while it is unstable if any are strictly greater than $1$. Accordingly, we find that across varying levels of fusion imprecision, there are natural boundaries dividing the $(\fusionrate,\evidencerate)$ parameter space into different stability regions. Specifically, we define $4$ possible regions in terms of different stability properties of the fixed points, as given in \Cref{table:region-class-table}. 

\begin{table}[h]
    \caption{$4$ possible stability regions.}
    \label{table:region-class-table}
    \centering
    \renewcommand{\arraystretch}{1.5}
    \begin{tabularx}{0.8\linewidth}{c>{\centering}X>{\centering}X>{\centering\arraybackslash}X}
        \toprule
        & $\beliefvector^f$& $\beliefvector^{h}$  & $\beliefvector^t$ \\
        \midrule
        $R_1$ & Unstable & Unstable  & Stable \\
        
        $R_2$ & Unstable & Stable  & Stable \\
        
        $R_3$  & Unstable & Unstable & Unstable \\
       
        $R_4$  & Stable & Stable & Stable \\
    
        \bottomrule
    \end{tabularx}
\end{table}

\begin{figure*}[t!]
\hspace{0.05\linewidth}
\centering
\includegraphics[width=0.75\linewidth]{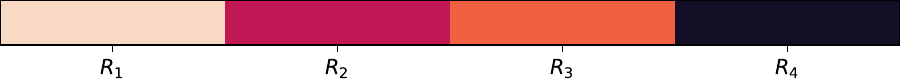}

\begin{minipage}{0.04\linewidth} 
\centering
\marginLeft{$\mathbf{\errorrate=0.4}$}
\end{minipage}
\begin{minipage}{0.95\linewidth}
\subfloat[$\gamma\in[0,\frac{1}{4})$]{
    \includegraphics[width = 0.23\linewidth]{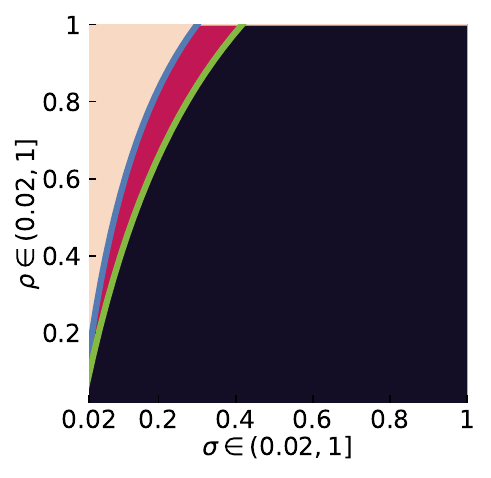}\label{fig:stability-regions-err-0.4-g-0}}
    \hfill
\subfloat[$\gamma\in[\frac{1}{4},\frac{1}{3})$]{\includegraphics[width=0.23\linewidth]{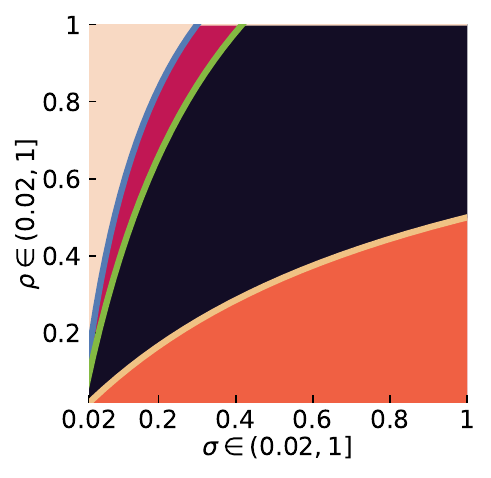}\label{fig:stability-regions-err-0.4-g-0.29}}
\hfill
\subfloat[$\gamma\in[\frac{1}{3},\frac{1}{2})$]{\includegraphics[width=0.23\linewidth]{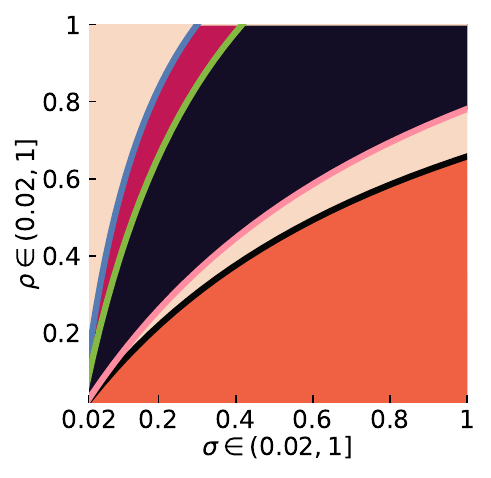}\label{fig:stability-regions-err-0.4-g-0.42}}
\hfill
\subfloat[$\gamma \geq\frac{1}{2}$]{\includegraphics[width=0.23\linewidth]{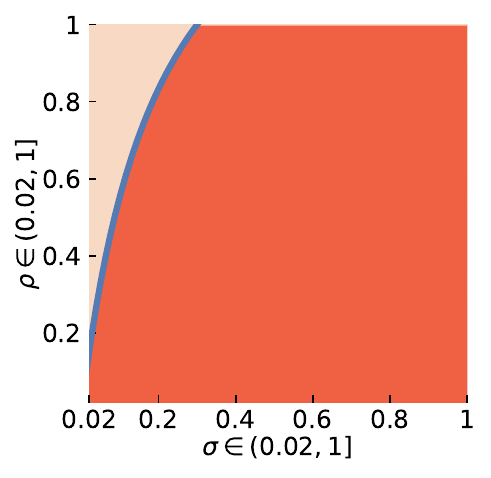}\label{fig:stability-regions-err-0.4-g-1}}
\hfill
% \subfloat[$\gamma=0.71$]{\includegraphics[width=0.45\linewidth]{diff_eq_sim/diff_eq_region_plots_g_0.71_h_0_err_0.4.pdf}}
% \subfloat[$\gamma=1$]{\includegraphics[width=0.45\linewidth]{diff_eq_sim/diff_eq_region_plots_g_1_h_0_err_0.4.pdf}}
\end{minipage}
\begin{minipage}{0.04\textwidth} 
\centering
\marginLeft{$\mathbf{\errorrate=0.3}$}
\end{minipage}
\begin{minipage}{0.95\textwidth} 
\subfloat[$\gamma\in[0,\frac{1}{4})$]{
    \includegraphics[width = 0.23\textwidth]{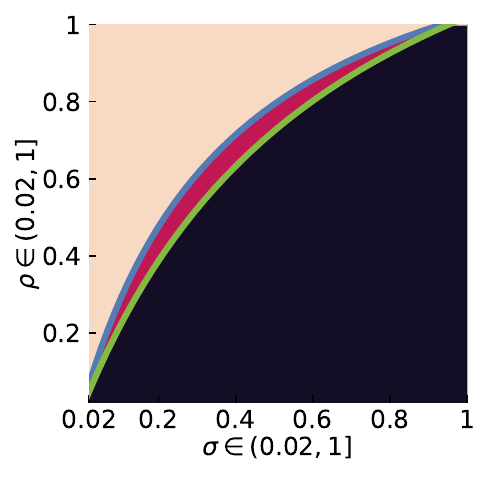}\label{fig:stability-regions-err-0.3-g-0}}
    \hfill
\subfloat[$\gamma\in[\frac{1}{4},\frac{1}{3})$]{\includegraphics[width=0.23\textwidth]{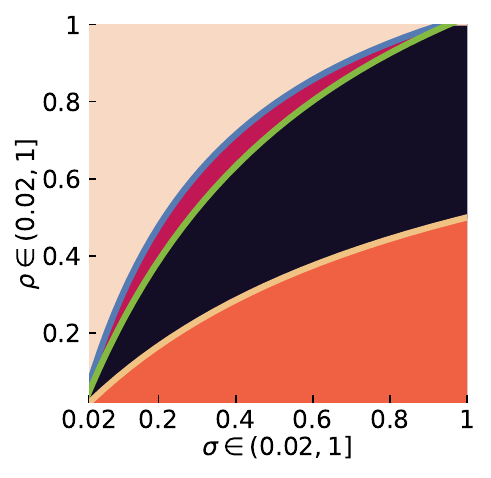}\label{fig:stability-regions-err-0.3-g-0.29}}
\hfill
\subfloat[$\gamma\in[\frac{1}{3},\frac{1}{2})$]{\includegraphics[width=0.23\textwidth]{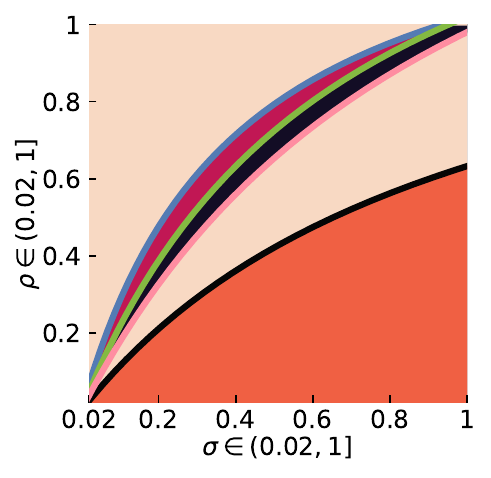}\label{fig:stability-regions-err-0.3-g-0.42}}
\hfill
\subfloat[$\gamma \geq\frac{1}{2}$]{\includegraphics[width=0.23\textwidth]{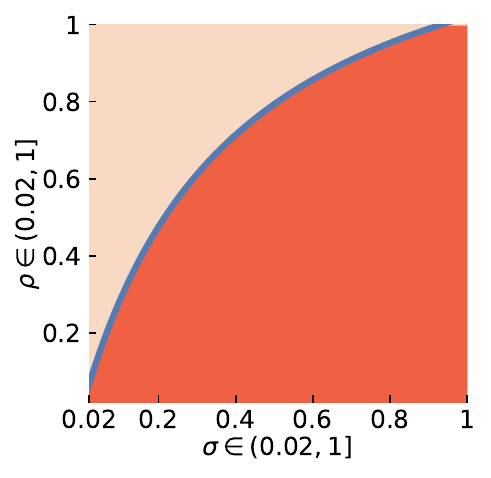}\label{fig:stability-regions-err-0.3-g-1}}
\hfill
\end{minipage}
    \caption{Heatmaps for the areas of $R_1$ to $R_4$ as defined in \Cref{table:region-class-table}, $\epsilon\in\{0.3,0.4\}$. $R_1$ represents the region where only $\beliefvector^t$ is stable, leading to optimal performance. $R_2$ is the region where only $\beliefvector^f$ is unstable, resulting in moderate performance. $R_3$ corresponds to the region where all fixed points are unstable, indicating that mixed performance should be expected. Finally, $R_4$ is the region where all fixed points are stable, which is associated with poor performance.}
    \label{fig:stability-regions-err}
\end{figure*}

In \Cref{fig:stability-regions-err} we show the stability regions of the $(\sigma,\rho)$ parameter space for the error rates $\errorrate=0.4$ and $\errorrate=0.3$ as the fusion imprecision $\gamma$ increases. In $R_1$ the incorrect fixed points, $\beliefvector^f$ and $\beliefvector^h$, are unstable equilibria while the correct fixed point $\beliefvector^t$ is stable. $R_2$ is the region in which both the correct and partially correct fixed points, $\beliefvector^t$ and $\beliefvector^h$, are stable,  while the false fixed point $\beliefvector^f$ is unstable. In $R_3$ all fixed points are unstable, while in $R_4$ they are all stable. In terms of learning performance for parameters within these different regions; In $R_1$ good performance can be expected under different initial conditions; In $R_2$ performance across initial conditions should be intermediate or good with stable convergence to at least a partially correct consensus; In $R_3$ the instability of all the fixed points may result in mixed performance; Finally, for fusion and evidence rates in $R_4$
we can expect poor performance under certain initial conditions and learning scenarios. In \Cref{fig:stability-regions-err}, the boundaries of the regions are defined by the following equations:

\begin{align}
    \text{Blue curve:} \quad & \evidencerate = \frac{\sigma}{(1 - \epsilon)(1 + \sigma - 2\epsilon)} \label{eq:blue_curve} \\
    \text{Green curve:} \quad & \evidencerate = \frac{\sigma}{1 + \sigma - 2\epsilon(1 + \sigma - \epsilon\sigma)} \label{eq:green_curve} \\
    \text{\revision{Pink} curve:} \quad & \evidencerate = \frac{\sigma}{\epsilon(1 + \sigma)(2 - \epsilon)} \label{eq:pink_curve} \\
    \text{Black curve:} \quad & \evidencerate = \frac{\sigma}{(1 + \sigma)(1 + \epsilon^2 - \epsilon)} \label{eq:white_curve} \\
    \text{Yellow curve:} \quad & \evidencerate = \frac{\sigma}{1 + \sigma} \label{eq:yellow_curve}
\end{align}
% In \Cref{fig:stability-regions-err-0.3-g-0.29,fig:stability-regions-err-0.4-g-0.29} we see that for a low level of imprecision ($\fthreshold\in[\frac{1}{4},\frac{1}{3})$) a mixed performance is expected under the white boundary compared with nonaccuracte consensus in the same region for precise fusion. This may be a benefit when the risk of incorrect consensus is high. 

 In general, the area of the $R_4$ region decreases as $\gamma$ increases. This typically results in an increase in $R_3$ due to the change in classification of $(\sigma,\rho)$ points from $R_4$ to $R_3$. Furthermore, there is an increase in the area of the $R_1$ region for some values of $\gamma$ and in particular for $\gamma \in [\frac{1}{3},\frac{1}{2})$, see \Cref{fig:stability-regions-err-0.3-g-0.42,fig:stability-regions-err-0.4-g-0.42}. In contrast, the $R_2$ region does not change significantly for most imprecision levels, with the exception of $\gamma \geq \frac{1}{2}$ were it is subsumed by $R_3$, see \Cref{fig:stability-regions-err-0.3-g-1,fig:stability-regions-err-0.4-g-1}.

% where the system is expected to be accurate above the blue boundary and exhibit mixed performance below it. 

Recalling that $R_1$ is characterised by robustly good performance, notice that the area of $R_1$ is maximal when $\gamma \in [\frac{1}{3},\frac{1}{2})$ corresponding to a medium level of imprecision,  as illustrated in \Cref{fig:stability-regions-err-0.4-g-0.42,fig:stability-regions-err-0.3-g-0.42}. In contrast the area of $R_4$, where consensus is likely to be highly dependent on initial conditions, is relatively low for $\gamma \in [\frac{1}{3},\frac{1}{2})$, especially for $\epsilon=0.3$ as shown in \Cref{fig:stability-regions-err-0.3-g-0.42}. Taken together these two factor suggest that  that this level of fusion imprecision may lead to optimal social learning performance by predominantly stabilising the correct consensus while minimising the stability of incorrect consensus states across the $(\evidencerate, \fusionrate)$ parameter space.

\begin{figure}[t]
    \centering
    % \begin{minipage}{0.05\linewidth} 
    %     \centering
    %     \marginLeft{\small Area/Benefits}
    % \end{minipage}
    \begin{minipage}{0.55\linewidth}
    \centering
        \figureAddLabel{0.9}{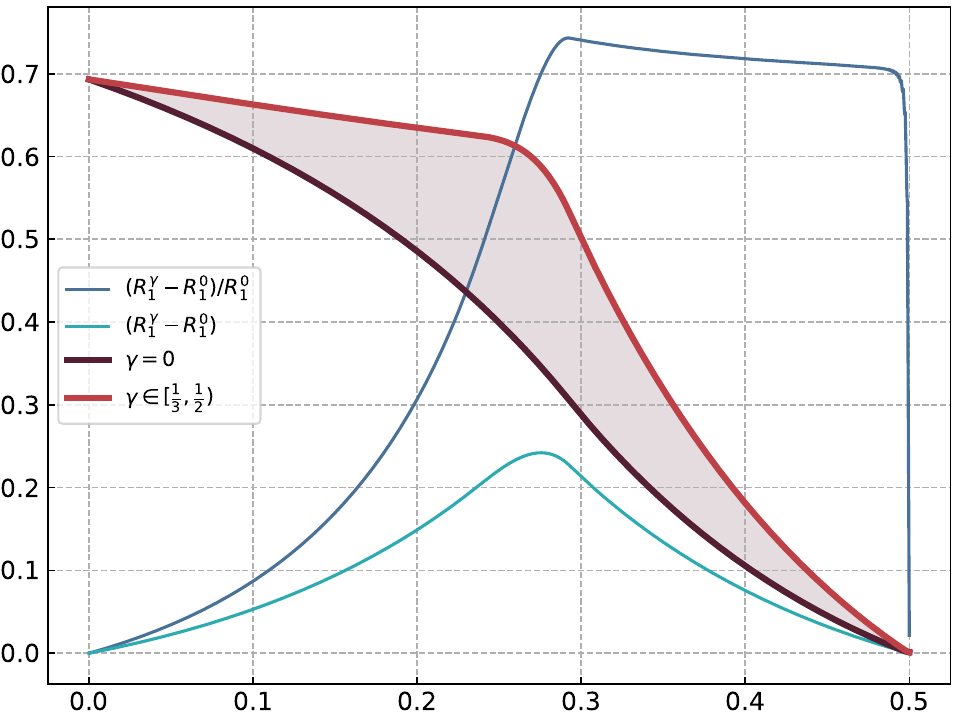}{\bottomText{\small Error rate $\errorrate$}}{\marginLeft{\small Area/Benefits}}[0.01][0.02]
    \end{minipage}
    \caption{Benefits of Imprecise Fusion Against Error Rates: Comparison of $R_1$ Area for Different Fusion Imprecision Levels. The blue and light blue(turquoise) lines represent the benefit ratio, $\frac{(R_1^\gamma - R_1^0)}{R_1^0}$, and the benefit $(R_1^\gamma - R_1^0)$ respectively. The red and purple curves show the area of $R_1$ for $\fthreshold \in \left[\frac{1}{3}, \frac{1}{2}\right)$ and $\fthreshold = 0$, respectively.}
    \label{fig:benefits-against-error}
\end{figure}

\Cref{fig:benefits-against-error}, shows the areas of $R_1$ for different levels of fusion imprecision against varying error rates, $\errorrate$. We let $R_1^\fthreshold$ denote the area of 
$R_1$ for fusion parameter $\fthreshold$, so that $R_1^0$ denotes the area of $R_1$ under precise fusion, and is shown as the red curve. The purple curve then corresponds to $R_1^\fthreshold$ for a moderate level of fusion imprecision where $\fthreshold \in \left[\frac{1}{3}, \frac{1}{2}\right)$. Hence, using $R_1^\fthreshold$ as a metric for robust good performance then this indicates that there are quantifiable benefits of imprecise fusion but that they vary across error rates $\errorrate\in\left[0,\ 0.5\right)$. In particular, for the noise-free case ($\errorrate=0$) we see that $R_1^0$ is over $70\%$ of the parameter space and that this area is the same for $\fthreshold \in \left[\frac{1}{3}, \frac{1}{2}\right)$. This suggests that there is no benefit from imprecise fusion when the error rate is $0$. However, in the presence of noisy evidence, for example at an error rate of $0.3$, $R_1^0$ shrinks significantly to approximately $30\%$ of the space. On the on the hand, $R_1^\fthreshold$ at a medium level of imprecision, covers a larger area of $50\%$ under the same noisy conditions. This increased coverage suggests that allowing for a certain degree of imprecision in the fusion process can enhance the system's robustness against the presence of evidential errors, and thereby supporting a more robust decision-making/social learning process.

A relative measure of robust performance, shown as the blue line in \Cref{fig:benefits-against-error}, is defined by the ratio $\frac{(R_1^\fthreshold - R_1^0)}{R_1^0}$, quantifying the increase in area of $R_1$ resulting from the level of imprecision $\fthreshold$, relative to the precise fusion case. For $\fthreshold \in \left[\frac{1}{3}, \frac{1}{2}\right)$ this relative measure increases with error rate up to a maximal value when $\errorrate \approx 0.28$ before a slight decline eventually followed by a sharp decrease as the error approaches $0.5$. This sharp decrease at higher error rates could be attributed to the overwhelming impact of noise on the evidence signal. The absolute benefit of imprecise fusion as given by $\delta{R_1}=(R_1^\fthreshold - R_1^0)$ is then shown by the light blue (turquoise) line, which also peaks at around $\errorrate=0.28$. The decline in absolute benefit beyond $\errorrate = 0.28$ signals a critical limit to the effectiveness of imprecise fusion under higher error conditions. This trend suggests that imprecise fusion can significantly enhance resilience and adaptability of the system under moderate error conditions. 

%such as autonomous vehicles, where precision and adaptability to environmental uncertainties are crucial. Understanding the relationship between error rates and fusion precision could also inform the development of protocols in distributed networks, where nodes must reliably aggregate data to achieve accurate overall system behavior.

\section{Simulation Results of the Difference Equations}\label{sec:difference-equation-simulations}

\begin{figure}[t!]
    \centering
    % \begin{minipage}{0.015\linewidth}
    %     \centering
    %     \marginLeft{\scriptsize Case 1}
    % \end{minipage}
     \begin{minipage}{0.5\linewidth}
     \centering
     \scriptsize{Color bar for Accuracy}\\
     \includegraphics[width=0.8\linewidth]{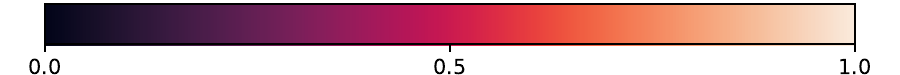}
        \vspace{1em} 
    \end{minipage}  \text{\textbf{\sffamily\bfseries\color{royalblue}{Case 1:}} population initialised at $\beliefvector_0^1=(0.01,0.01,0.01,0.97,\mathbf{0})$ }
    \begin{minipage}{1\linewidth}
    {\subfloat[$\gamma=0$]{\includegraphics[width=0.136\linewidth]{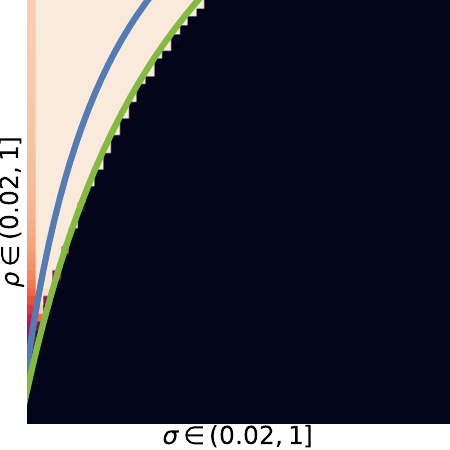}\label{fig:diff-eq-sim-1-1-1-97-0}}}
    \hfill
    {\subfloat[$\gamma=\frac{1}{4}$]{\includegraphics[width=0.136\linewidth]{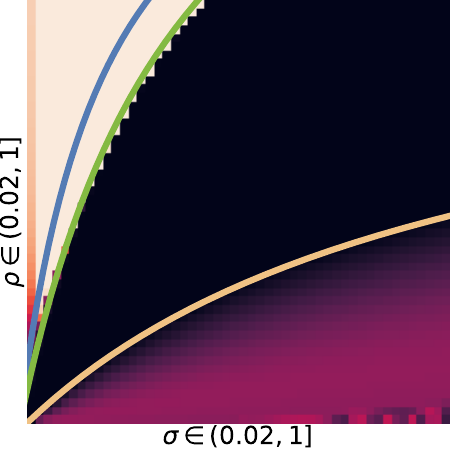}\label{fig:diff-eq-sim-1-1-1-97-0.29}}}
    \hfill
    {\subfloat[$\gamma=\frac{1}{3}$]{\includegraphics[width=0.136\linewidth]{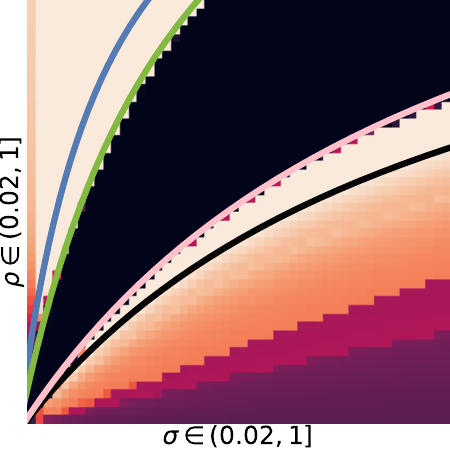}\label{fig:diff-eq-sim-1-1-1-97-0.42}}}
    \hfill
    {\subfloat[$\gamma=\frac{1}{2}$]{\includegraphics[width=0.136\linewidth]{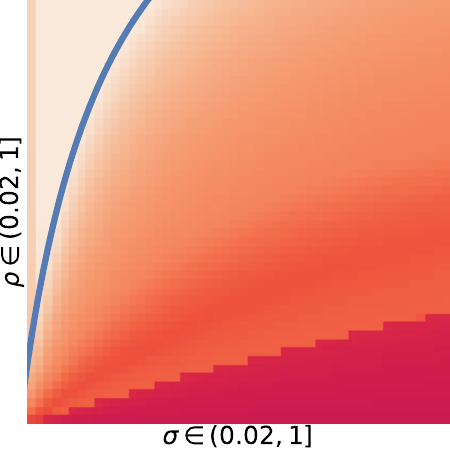}\label{fig:diff-eq-sim-1-1-1-97-0.58}}}
    \hfill
    {\subfloat[$\gamma=\frac{2}{3}$]{\includegraphics[width=0.136\linewidth]{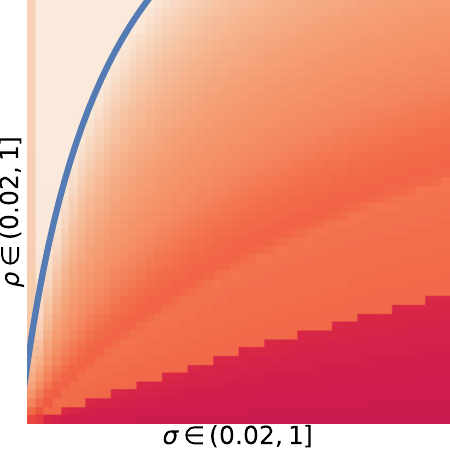}\label{fig:diff-eq-sim-1-1-1-97-0.71}}}
    \hfill
    {\subfloat[$\gamma=\frac{3}{4}$]{\includegraphics[width=0.136\linewidth]{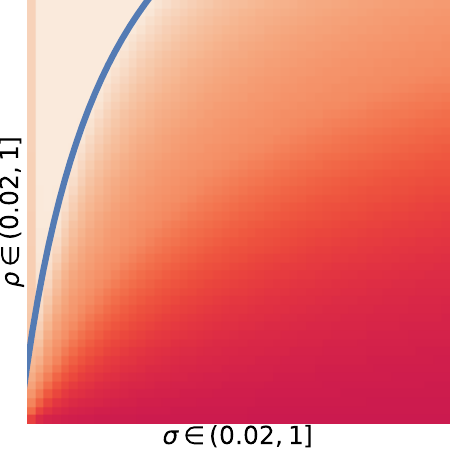}\label{fig:diff-eq-sim-1-1-1-97-0.8}}}
    \hfill
    {\subfloat[$\gamma=1$]{\includegraphics[width=0.136\linewidth]{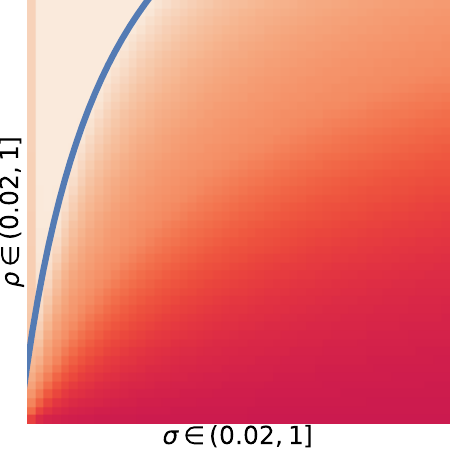}\label{fig:diff-eq-sim-1-1-1-97-1}}}
        \vspace{1em}
    \end{minipage}
    % \begin{minipage}{0.015\linewidth}
    % \centering
    % \marginLeft{\small Case 2}
    % \end{minipage}    
   \text{\textbf{\sffamily\bfseries\color{royalblue}{Case 2:}} population initialised at $\beliefvector_0^2=(0.01,0.01,0.97,0.01,\mathbf{0})$}
      
    \begin{minipage}{0.999\linewidth}
    {\subfloat[$\gamma=0$] {\includegraphics[width=0.136\linewidth]{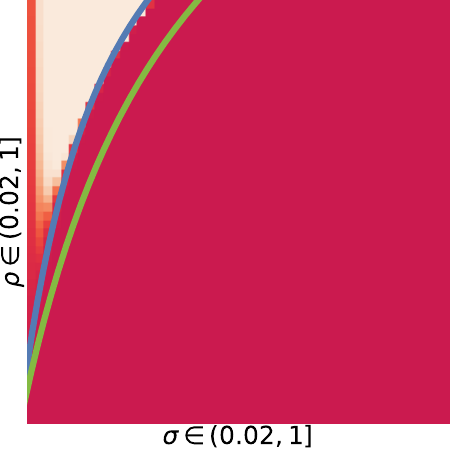}\label{fig:diff-eq-sim-1-1-97-1-0}}}
    \hfill
    {\subfloat[$\gamma=\frac{1}{4}$]{\includegraphics[width=0.136\linewidth]{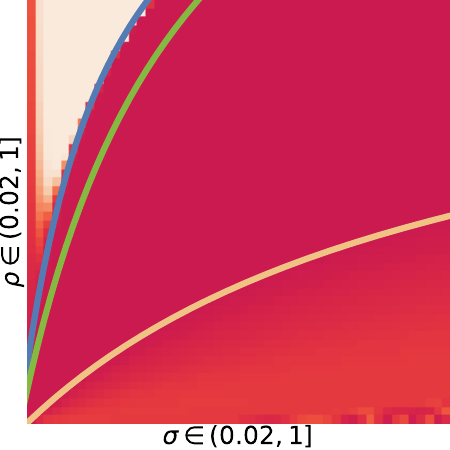}\label{fig:diff-eq-sim-1-1-97-1-0.29}}}  
    \hfill
    {\subfloat[$\gamma=\frac{1}{3}$]{\includegraphics[width=0.136\linewidth]{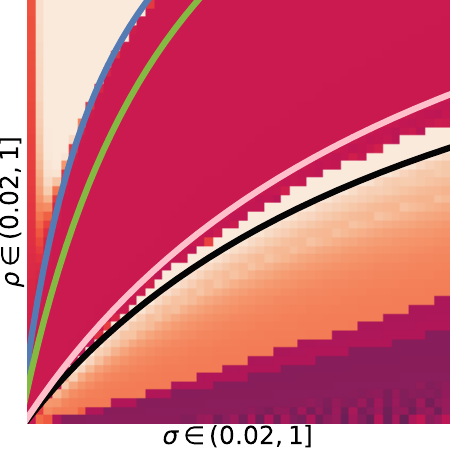}\label{fig:diff-eq-sim-1-1-97-1-0.42}}}
    \hfill
    {\subfloat[$\gamma=\frac{1}{2}$]{\includegraphics[width=0.136\linewidth]{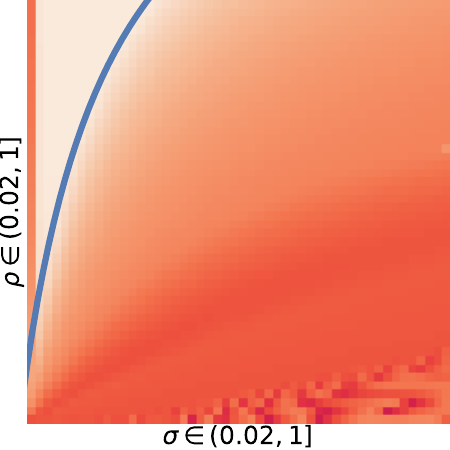}\label{fig:diff-eq-sim-1-1-97-1-0.58}}}
    \hfill
    {\subfloat[$\gamma=\frac{2}{3}$]{\includegraphics[width=0.136\linewidth]{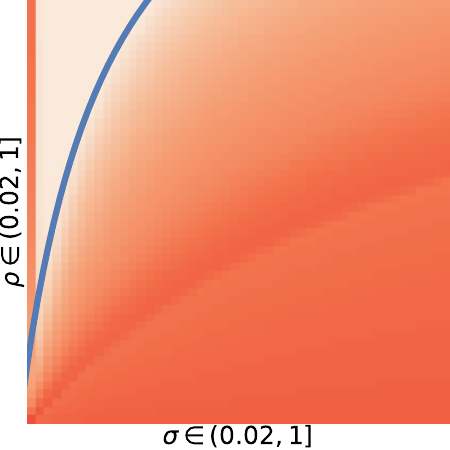}\label{fig:diff-eq-sim-1-1-97-1-0.71}}}
    \hfill
    {\subfloat[$\gamma=\frac{3}{4}$]{\includegraphics[width=0.136\linewidth]{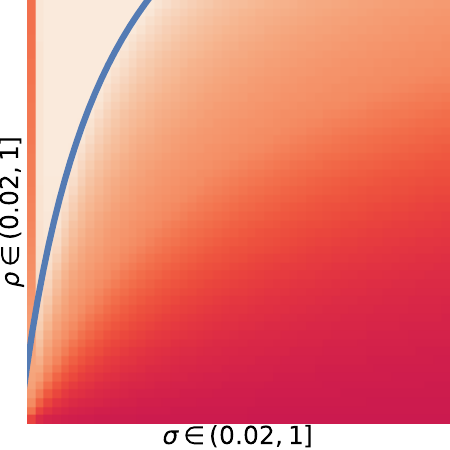}\label{fig:diff-eq-sim-1-1-97-1-0.8}}}
    \hfill   
    {\subfloat[$\gamma=1$]{\includegraphics[width=0.136\linewidth]{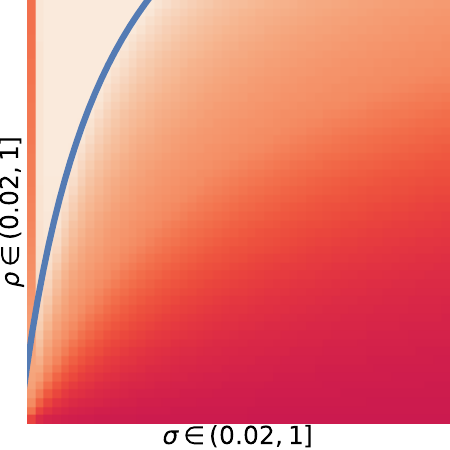}\label{fig:diff-eq-sim-1-1-97-1-1}}}
    \vspace{1em}
    \end{minipage}
    % \begin{minipage}{0.015\linewidth}
    % \centering
    % \marginLeft{\small Case 3}
    % \end{minipage}

\text{\textbf{\sffamily\bfseries\color{royalblue}{Case 3:}} population initialised at $\beliefvector_0^3=(0.97,0.01,0.01,0.01,\mathbf{0})$}    
    \begin{minipage}{0.999\linewidth}
    {\subfloat[$\gamma=0$]{\includegraphics[width=0.136\linewidth]{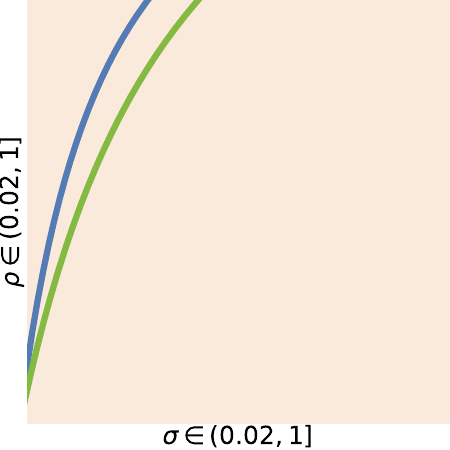}\label{fig:diff-eq-sim-97-1-1-1-0}}}
    \hfill
    {\subfloat[$\gamma=\frac{1}{4}$]{\includegraphics[width=0.136\linewidth]{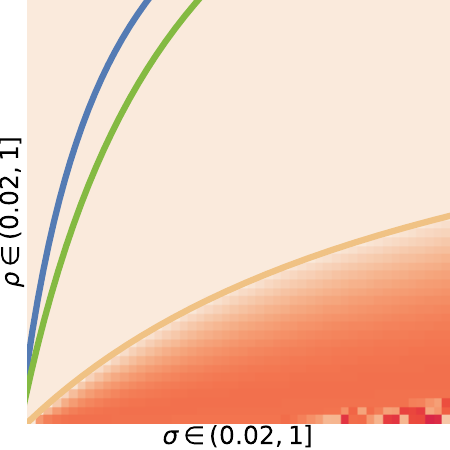}\label{fig:diff-eq-sim-97-1-1-1-0.29}}}
    \hfill
    {\subfloat[$\gamma=\frac{1}{3}$]{\includegraphics[width=0.136\linewidth]{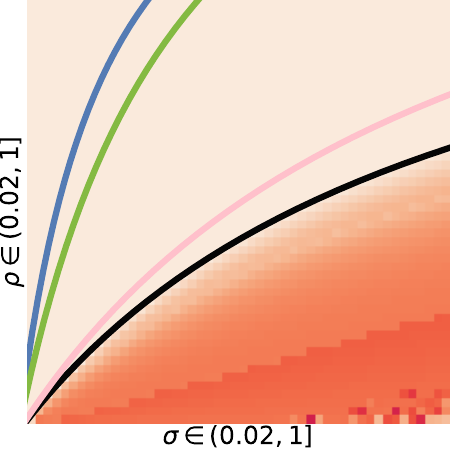}\label{fig:diff-eq-sim-97-1-1-1-0.42}}}
    \hfill
    {\subfloat[$\gamma=\frac{1}{2}$]{\includegraphics[width=0.136\linewidth]{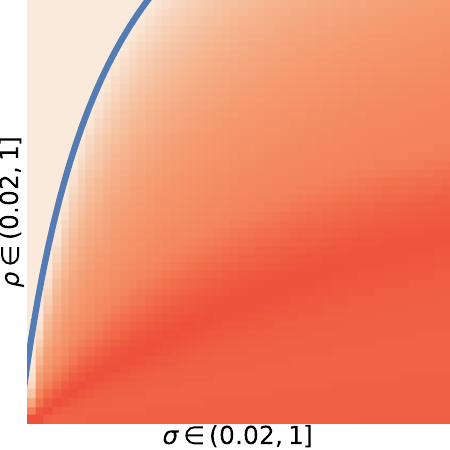}\label{fig:diff-eq-sim-97-1-1-1-0.58}}}
    \hfill
    {\subfloat[$\gamma=\frac{2}{3}$]{\includegraphics[width=0.136\linewidth]{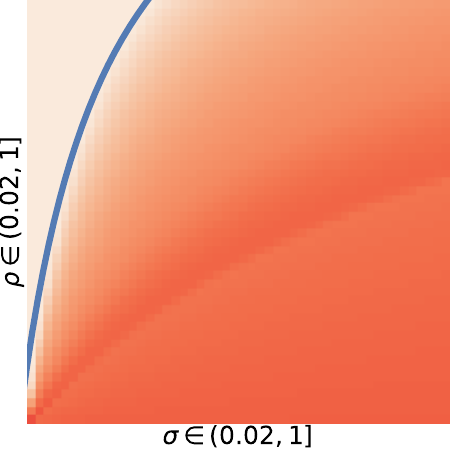}\label{fig:diff-eq-sim-97-1-1-1-0.71}}}
    \hfill{\subfloat[$\gamma=\frac{3}{4}$]{\includegraphics[width=0.136\linewidth]{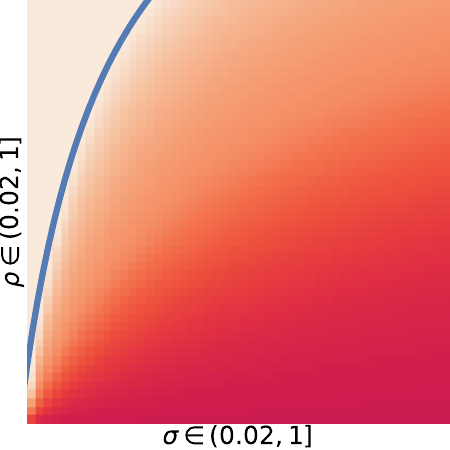}\label{fig:diff-eq-sim-97-1-1-1-0.8}}}
    \hfill
    {\subfloat[$\gamma=1$]{\includegraphics[width=0.136\linewidth]{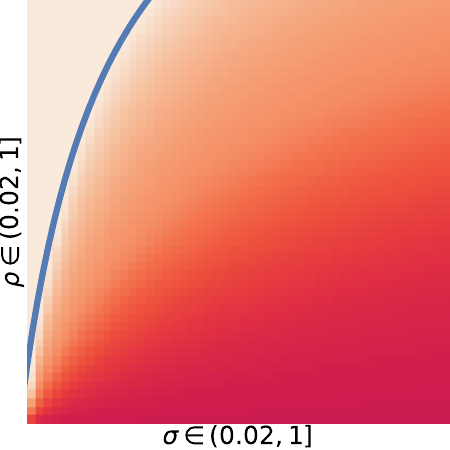}\label{fig:diff-eq-sim-97-1-1-1-1}}}
    \end{minipage}
    \caption{Heat maps showing the proportion of correct beliefs with different combinations of fusion rates ($\fusionrate$) and evidence rates ($\evidencerate$), for the various fusion imprecision initialised at 3 different conditions. The results are at $t = 1500$ and $\epsilon = 0.4$. The stability boundaries of each $\gamma$ are also marked. \revision{For Cases 1 \& 2, imprecise fusion improves learning accuracy when most agents initially hold incorrect beliefs. The strongest benefits appear in \cref{fig:diff-eq-sim-1-1-1-97-0.42,fig:diff-eq-sim-1-1-97-1-0.42}.}}\label{fig:diff-eq-sim}
\end{figure}

In this section we present simulation results for different levels of fusion imprecision under different initial conditions. We consider $3$ different scenarios characterised by distinct initial conditions: {\bf Case $1$}: a population of agents with initial beliefs such that $97\%$ of agents begin the learning process committed to the incorrect belief, and $1\%$ committed to each of the other $3$ precise beliefs, i.e. $\beliefvector_0^1=(0.01,0.01,0.01,0.97,\mathbf{0})$. The simulation results for {\bf Case $1$} are shown in \Crefrange{fig:diff-eq-sim-1-1-1-97-0}{fig:diff-eq-sim-1-1-1-97-1}; {\bf Case $2$}: a population of agents with initial beliefs such that $97\%$ of the agents begin the learning process committed to a partially correct belief and with $1\%$ committed to each of the other $3$ precise beliefs, i.e. without loss of generality we take $\beliefvector_0^2=(0.01,0.01, 0.97,0.01,\mathbf{0})$. The simulation results for {\bf Case $2$} are shown in \Crefrange{fig:diff-eq-sim-1-1-1-97-0}{fig:diff-eq-sim-1-1-1-97-1}; {\bf Case $3$}: a population of agents with initial beliefs such that $97\%$ of the agents begin the learning process committed to the correct belief and with $1\%$ committed to each of the other $3$ precise beliefs, i.e. $\beliefvector_0^3=(0.97,0.01,0.01,0.01,\mathbf{0})$. The various parameter settings are listed in \Cref{table:parameters} for better clarity. The simulation results for {\bf Case $3$} are shown in  \Crefrange{fig:diff-eq-sim-97-1-1-1-0}{fig:diff-eq-sim-97-1-1-1-1}.

Performance is evaluated for these three scenarios using the average Hamming distance $H$ from the agents' beliefs to the true state $s^*$. Furthermore, without loss of generality, we assume that $s^*$ is such that $s^*(p_i)=1$ for $i=1, \ldots, n$. In this context, the Hamming distance between states is defined as follows:  Let $s_1= \langle s_1(p_1),\dots,s_1(p_n) \rangle$ and $s_2= \langle s_2(p_1),\dots,s_2(p_n) \rangle$ be two states, then the Hamming distance between them is given by:

\begin{gather}\label{eq:hamming-distance}
H(s_1,s_2)=\sum_{i=1}^n \lvert s_1(p_i)-s_2(p_i) \lvert {.}
\end{gather}

We then extend this to give a normalised Hamming distance between a belief $B \subseteq \mathbb{S}$ and the true state of the world $s^*$ as follows:
\begin{gather}
H(B,s^*)=\frac{1}{\lvert B \lvert} \frac{1}{n} \sum_{s \in B} ~H(s,s^*) {.}
\label{eq:avg-hamming-distance}
\end{gather}

Furthermore, we evaluate the performance at the population level, $\alpha$, as the average Hamming distance between the population of agents $\mathcal{A}$ of size $k$, and $s^*$, re-scaled as an accuracy measure such that:

\begin{gather}
    \accuracy(\beliefvector) = 1 - \beliefvector\cdot\mathbf{H}^{\intercal} ,
\end{gather}
where $\mathbf{H} = \left[ H(B_i, s^*) \right]_{i=1}^{15}$.  
More specifically, for the indexing of beliefs given in \Cref{tab:belief-notations}, 
\begin{gather*}
    \mathbf{H} =\begin{smallmatrix}
    [0 & \frac{1}{2}& \frac{1}{2} & 1 & \frac{1}{4}& \frac{1}{4} & \frac{1}{2}& \frac{1}{2} & \frac{3}{4} &\frac{3}{4} & \frac{1}{3} & \frac{1}{2} & \frac{1}{2} & \frac{2}{3} & \frac{1}{2}]
\end{smallmatrix}
\end{gather*}
In other words, if all agents reach consensus at $\truestate$, then the learning accuracy is $\accuracy(\beliefvector^t)=1$, whereas for the other two fixed points it is  $\accuracy(\beliefvector^f)=0$ and $\accuracy(\beliefvector^h)=\frac{1}{2}$ respectively.

\Cref{fig:diff-eq-sim} shows that for all levels of imprecision and initial conditions, the learning results are usually accurate when the fusion rate $\fusionrate$ is low while the evidence rate $\evidencerate$ is high, even in {\bf Case 1} where $97\%$ of the agents are initially committed to the incorrect beliefs. For the other regions of the parameter space, when precise fusion is applied  the agents typically reach consensus on the belief to which most of the population were initially committed. This is not problematic for {\bf Case 3}, when most agents are initially committed to the true belief, and in fact in that case imprecise fusion can have a detrimental effect on performance by inhibiting convergence to a shared precise belief. However, when most agents are initialised with an incorrect belief, as in {\bf Case 1} and {\bf Case 2}, we can see the benefits of imprecise fusion. 

More specifically, for both {\bf Case 1} and {\bf Case 2} we see from \Crefrange{fig:diff-eq-sim-1-1-1-97-0}{fig:diff-eq-sim-1-1-97-1-1} that imprecise fusion can improve learning accuracy when most agents are initially committed to an incorrect belief. The most significant benefits of imprecise fusion can be seen in \Cref{fig:diff-eq-sim-1-1-1-97-0.42,fig:diff-eq-sim-1-1-97-1-0.42}, where the fusion threshold $\fthreshold\in[\frac{1}{3},\frac{1}{2})$, and where very high accuracy $\accuracy \approx 1$ is reach for more $(\sigma,\rho)$ points. This is consistent with the fixed point analysis in \Cref{sec:fixed-point-analysis} and in particular the stability regions shown  in \Cref{fig:stability-regions-err}. For instance, for both cases performance in region $R_1$ is good across the different imprecision levels. Also, in region $R_4$ accuracy corresponds to a value consistent with the initial conditions, suggesting convergence to the dominant initial belief in both {\bf Case 1} and {\bf Case 2}. As is consistent with the instability of all fixed points, in region $R_3$ we see a mixed level of accuracy, which tends to be higher when the evidence rate is higher, and which is often consistent with a failure to reach consensus on a precise belief, especially for higher levels of imprecise fusion. Consider, for example, \Cref{fig:diff-eq-sim-1-1-1-97-0.42} corresponding to {\bf Case 1} accuracy when $\gamma=\frac{1}{3}$. The relevant stability regions are then shown in \Cref{fig:stability-regions-err-0.4-g-0.42} and it is clear that in this case there is a strong correlation between accuracy and stability region, with high accuracy in $R_1$ and $R_2$, low accuracy in $R_4$ and variable accuracy in $R_3$ which is higher for those $R_3$ point with higher evidence rates.

\begin{table}[t]
\centering
\caption{Overview of model parameters}
\label{table:parameters}
\setlength{\tabcolsep}{10pt} % 控制列间距
\renewcommand{\arraystretch}{1.3} % 控制行高
\begin{tabular}{>{\raggedright\arraybackslash}p{5cm} >{\raggedright\arraybackslash}p{6cm}}
\toprule
\rowcolor[HTML]{EFEFEF} % 设置表头背景色
\textbf{Parameter} & \textbf{Values} \\
\midrule
\rowcolor[HTML]{F7F7F7} % 设置背景色
Population size & $k\in\{100,200,400\}$ agents \\
Maximum simulation time steps & 20,000 iterations \\
\rowcolor[HTML]{F7F7F7}
Evidence rate & $\evidencerate\in(0,1)$ \\
Fusion rate  & $\fusionrate\in(0,1)$ \\
\rowcolor[HTML]{F7F7F7}
Error rate  & $\errorrate\in\{0.2,0.3,0.4\}$ \\
Language size & $n=2$ \\
\rowcolor[HTML]{F7F7F7}
Number of possible states & $4$ \\
Number of possible agent's belief & 15 \\
\rowcolor[HTML]{F7F7F7}
Initial biased belief $\beliefvector_0$& Case 1: (0.05,0.01,0.01,0.97,$\mathbf{0}$) \newline Case 2: (0.01,0.01,0.97,0.01,$\mathbf{0}$)  \newline Case 3: (0.97,0.01,0.01,0.01,$\mathbf{0}$)\\
& Case 4: (0.05,0.05,0.05,0.85,$\mathbf{0}$) \newline Case 5: (0.05,0.05,0.85,0.05,$\mathbf{0}$)  \newline Case 6: (0.85,0.05,0.05,0.05,$\mathbf{0}$)\\ 
\rowcolor[HTML]{F7F7F7} &
Case 7: (0.1,0.1,0.1,0.7,$\mathbf{0}$) \newline Case 8: (0.1,0.1,0.7,0.1,$\mathbf{0}$)  \newline Case 9: (0.7,0.1,0.1,0.1,$\mathbf{0}$)
\\
\bottomrule
\end{tabular}
\end{table}

\begin{figure*}[t!]
\centering
\begin{minipage}{0.87\linewidth}
\begin{minipage}{1\textwidth} 
\centering
\subfloat[$\gamma=0$]{\figureAddLabel{0.315}{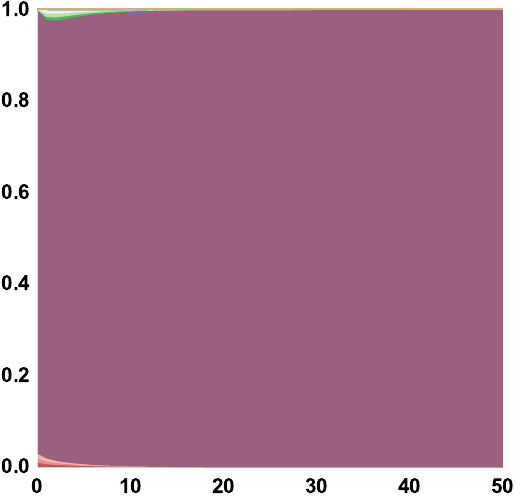}{\bottomText{\scriptsize{Time Steps $t$}}}{\marginLeft{\scriptsize{Proportion}}}\label{fig:sankey-err-0.4-g-0}}
\hfill
\subfloat[$\gamma=\frac{1}{3}$]{\figureAddLabel{0.315}{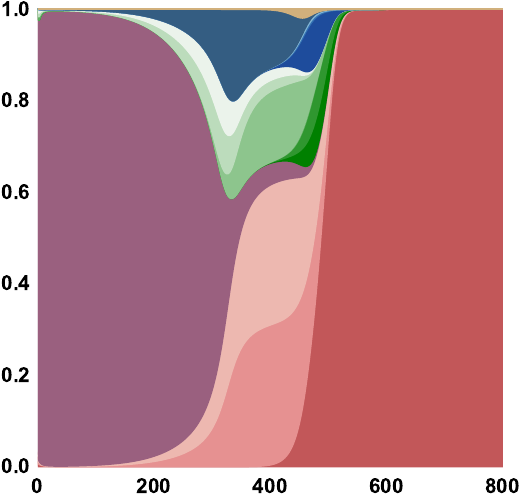}{\bottomText{\scriptsize{Time Steps $t$}}}{}\label{fig:sankey-err-0.4-g-0.42}}
\hfill
\subfloat[$\gamma=\frac{3}{4}$]{\figureAddLabel{0.315}{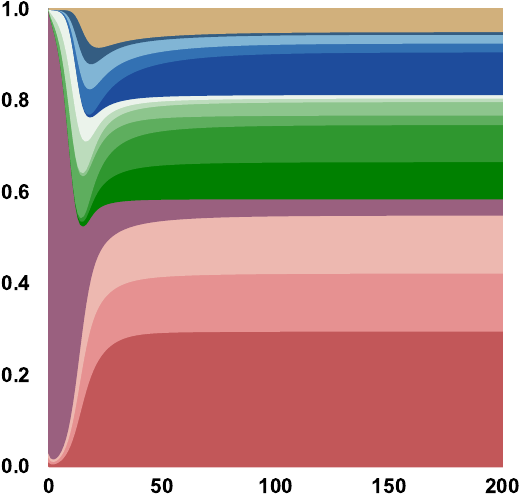}{\bottomText{\scriptsize{Time Steps $t$}}}{}\label{fig:sankey-err-0.4-g-0.8}}
\end{minipage}
% \begin{minipage}{0.04\textwidth} 
% \centering
% \marginLeft{\scriptsize{Entropy/Precision}}
% \end{minipage}
\begin{minipage}{1\textwidth} 
\subfloat[$\gamma=0$]{\figureAddLabel{0.315}{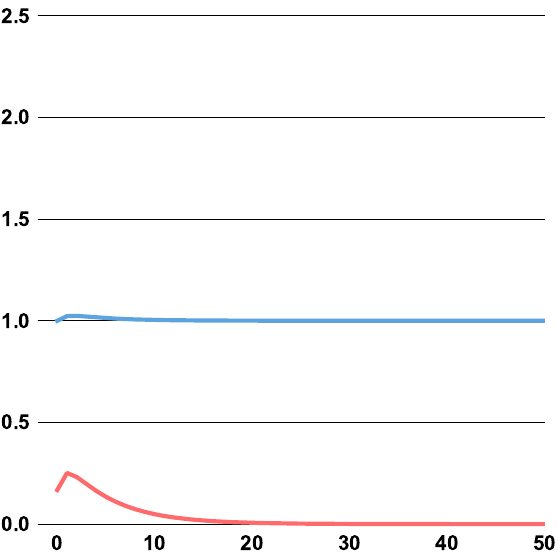}{\bottomText{\scriptsize{Time Steps $t$}}}{\marginLeft{\scriptsize{Entropy/Imprecision}}}\label{fig:entropy-err-0.4-g-0}}
\hfill
\subfloat[$\gamma=\frac{1}{3}$]{\figureAddLabel{0.315}{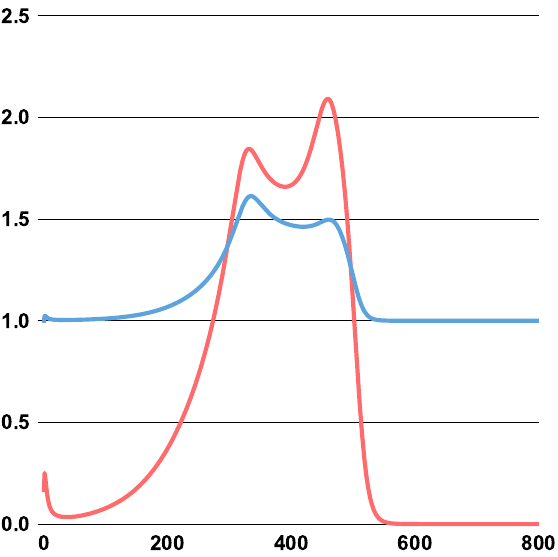}{\bottomText{\scriptsize{Time Steps $t$}}}{}\label{fig:entropy-err-0.4-g-0.42}}
\hfill
\subfloat[$\gamma=\frac{3}{4}$]{\figureAddLabel{0.315}{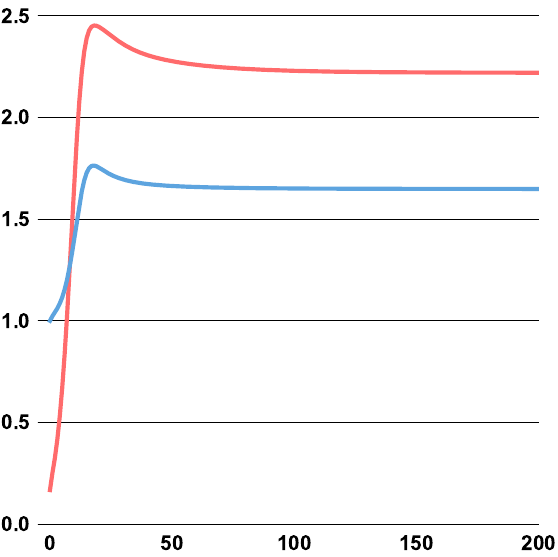}{\bottomText{\scriptsize{Time Steps $t$}}}{}\label{fig:entropy-err-0.4-g-0.8}}
\end{minipage}
\end{minipage}
\begin{minipage}{0.12\linewidth}
    \includegraphics[width=\linewidth]{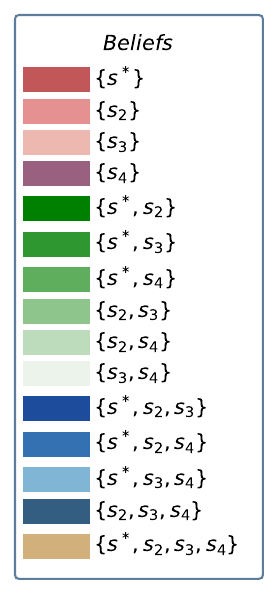}
    \includegraphics[width=\linewidth]{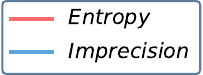}
\end{minipage}
\caption{Sankey plots (\ref{fig:sankey-err-0.4-g-0} to \ref{fig:sankey-err-0.4-g-0.8}) for the proportions of  beliefs $B_1$ to $B_{15}$ against time step $t$. Also, line plots of entropy and imprecision against time step $t$ (\ref{fig:entropy-err-0.4-g-0} to \ref{fig:entropy-err-0.4-g-0.8}). Results are for {\bf Case 1}, with $\evidencerate=0.5$, $\fusionrate=0.5$, and $\epsilon=0.4$. This combination of $(\evidencerate,\fusionrate)$ belongs to regions $R_4$, $R_1$, and $R_3$ for $\fthreshold=0, \ \frac{1}{3}$and $\frac{3}{4}$ respectively.}
    \label{fig:time-series}
\end{figure*}

\begin{figure}[htb]
\centering
\begin{minipage}{0.7\linewidth}
\begin{minipage}{1\linewidth} 
\centering
% \subfloat[$\gamma=0$]{\figureAddLabel{0.45}{time_series/sankey_plot_evi_0.5_f_0.5_g_0_h_0_err_0.4_initial__97_01_01_01.pdf}{\bottomText{\scriptsize{Time Steps $t$}}}{\marginLeft{\scriptsize{Proportion}}}\label{fig:sankey-from-correct-err-0.4-g-0}}
% \hfill
% \subfloat[$\gamma=0.42$]{\figureAddLabel{0.315}{time_series/sankey_plot_evi_0.5_f_0.5_g_0.42_h_0_err_0.4_initial__97_01_01_01.pdf}{\bottomText{\scriptsize{Time Steps $t$}}}{}\label{fig:time-series-from-correct-err-0.4-g-0.42}}
\subfloat[$\gamma=\frac{3}{4}$]{\figureAddLabel{0.45}{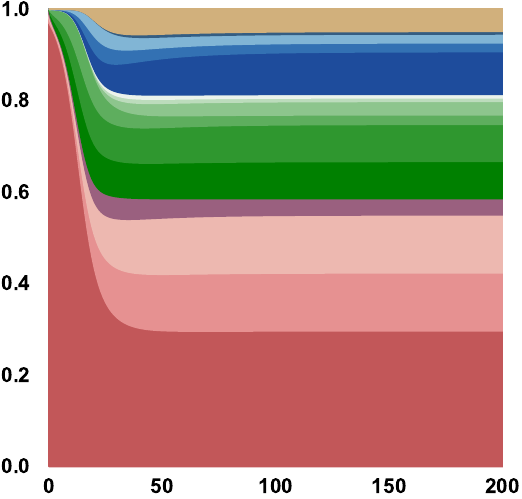}{\bottomText{\scriptsize{Time Steps $t$}}}{\marginLeft{\scriptsize{Proportion}}}\label{fig:sankey-from-correct-err-0.4-g-0.8}}
% \end{minipage}
% \begin{minipage}{0.04\textwidth} 
% \centering
% \marginLeft{\scriptsize{Entropy/Precision}}
% \end{minipage}
% \begin{minipage}{0.4\textwidth} 
% \subfloat[$\gamma=0$]{\figureAddLabel{0.45}{time_series/entropy_plot_evi_0.5_f_0.5_g_0_h_0_err_0.4_initial__97_01_01_01.pdf}{\bottomText{\scriptsize{Time Steps $t$}}}{\marginLeft{\scriptsize{Entropy/Imprecision}}}\label{fig:entropy-from-correct-err-0.4-g-0}}
% \hfill
% \subfloat[$\gamma=0.42$]{\figureAddLabel{0.315}{time_series/entropy_plot_evi_0.5_f_0.5_g_0.42_h_0_err_0.4_initial__97_01_01_01.pdf}{\bottomText{\scriptsize{Time Steps $t$}}}{}\label{fig:entropy-from-correct-err-0.4-g-0.42}}
\hfill
\subfloat[$\gamma=\frac{3}{4}$]{\figureAddLabel{0.44}{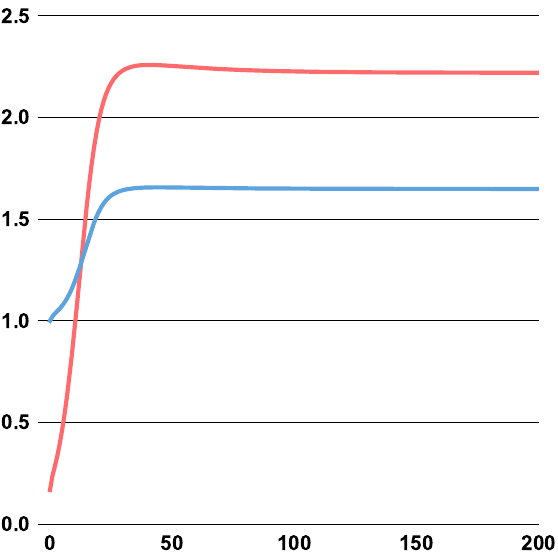}{\bottomText{\scriptsize{Time Steps $t$}}}{\marginLeft{\scriptsize{Proportion}}}\label{fig:entropy-from-correct-err-0.4-g-0.8}}
\end{minipage}

\end{minipage}
\begin{minipage}{0.2\linewidth}
    \includegraphics[width=0.8\linewidth]{time_series/color_legend.pdf}
    \includegraphics[width=0.8\linewidth]{time_series/color_legend_entropy_plot.pdf}
\end{minipage}
\caption{Sankey plots (\ref{fig:sankey-from-correct-err-0.4-g-0.8}) for the proportions of beliefs $B_1$ to $B_{15}$ against time step $t$.  Also, line plots of entropy and imprecision against times step $t$ ( \ref{fig:entropy-from-correct-err-0.4-g-0.8}). Results are for {\bf Case 3} with  $\evidencerate=0.5$, $\fusionrate=0.5$, and $\epsilon=0.4$. This combination of $(\evidencerate,\fusionrate)$ belongs to and $R_3$ for $\frac{3}{4}$, respectively \label{fig:time-series-from-correct}}
\end{figure}

%\section{Population Dynamics}\label{timeseries}
\revision{We can gain further insight into the effect of imprecise fusion in the proposed model by investigating how the proportions of different beliefs in the population vary with time under specified social learning conditions and for varying levels of imprecision. Specially, using the different equation model we can consider how the various proportions $\mathbf{P}_t(i)$ change during the social learning process, as $t$ ranges from $0$ to $800$ time steps, for different initial conditions $\mathbf{P}_0$ and fusion imprecision levels $\gamma$. \Crefrange{fig:sankey-err-0.4-g-0}{fig:sankey-err-0.4-g-0.8} are sankey plots showing the proportions of different beliefs against time for imprecision levels $\gamma=0$ (precise fusion), $\gamma=\frac{1}{3}$ (moderately imprecise fusion) and $\gamma=\frac{3}{4}$ (highly imprecise fusion)  respectively. These results are for  {\bf Case 1} where the initial beliefs are highly biased towards the completely incorrect belief $B=\{s_4\}$. The results are for equal rates of evidential updating and fusion where $\sigma=\rho=0.5$ corresponding to the centre of the $(\sigma,\rho)$ parameter space as shown in \Cref{fig:diff-eq-sim}.
For precise fusion as shown in \Cref{fig:sankey-err-0.4-g-0}, it is clear that despite the early introduction of a small proportion of alternative beliefs the population quickly convergences on the false belief towards which the population is initially highly biased. In contrast, for moderately imprecise fusion as shown in \Cref{fig:sankey-err-0.4-g-0.42}, the diversity of opinions present in the population increases significantly during an intermediate learning period before the population reaches consensus on the true state, i.e. $B=\{s^*\}$. The intermediate beliefs are mostly imprecise but also consistent with the true state, i.e. $s^* \in B$. We might therefore conjecture that the emergence of such beliefs provides an incentive for agents to search for evidence much of which will then support the true state. However, if fusion is too imprecise as in the case of \Cref{fig:sankey-err-0.4-g-0.8}, then the initial growth of imprecise opinions stabilises resulting in a diverse population without consensus. Interestingly, this is also true of {\bf Case 3} in which agents are initially highly biased towards the true state. In this case, while both precise and moderately imprecise fusion results in rapid convergence to population wide consensus on the true state, highly imprecise fusion still leads to a persistently diverse population holding a variety of imprecise opinions (see \Cref{fig:sankey-from-correct-err-0.4-g-0.8}). }

\revision{We can quantify these dynamic effects of imprecise fusion on the agent population using two measures. As a measure of population diversity we can use the entropy of the current proportions $\mathbf{P}_t$ as follows: 
\begin{gather*}
H_t = \sum_{i} -\beliefvector_t(i)\cdot\log_2\beliefvector_t(i)
\end{gather*}
As a measure of imprecision we propose the expected cardinality of the population's beliefs at any given time as follows:
\begin{gather*}
    I_t = \sum_{i}\beliefvector_t(i)\cdot|B_i|
\end{gather*}
\Crefrange{fig:entropy-err-0.4-g-0}{fig:entropy-err-0.4-g-0.8} show the population entropy (red line) and imprecision (blue line) plotted against time in {\bf Case 1} for the three levels of imprecise fusion. For precise fusion (\Cref{fig:entropy-err-0.4-g-0}) both entropy and imprecision remain low throughout the simulation as is consistent with very little deviation from the totally incorrect belief $B=\{s_4\}$ to which there is an initial strong bias. For moderately imprecise fusion \Cref{fig:entropy-err-0.4-g-0.42} shows that both imprecision and entropy increase during an initial learning period before decrease as the population reaches consensus on the true belief, i.e. $B=\{s^*\}$. In contrast, for highly imprecise fusion as shown in \Cref{fig:entropy-err-0.4-g-0.8}, an early peak is followed by convergence to a high value of both measures indicating convergence to a diverse set of imprecise opinions. Indeed similar behaviour is also found for {\bf Case 3} as shown in \Cref{fig:entropy-from-correct-err-0.4-g-0.8}.}

%\begin{figure}[htb]
%\centering
%\begin{minipage}{0.7\linewidth}
%\begin{minipage}{1\textwidth} 
%\centering
%\subfloat[$\gamma=\frac{1}{3},\evidencerate=0.1,\fusionrate=0.9$]{\figureAddLabel{0.45}{time_series/sankey_plot_evi_0.1_f_0.9_g_0.42_h_0_err_0.4_initial__01_01_97_01.pdf}{\bottomText{\scriptsize{Time Steps $t$}}}{\marginLeft{\scriptsize{Proportion}}}\label{fig:sankey-from-half-correct-err-0.4-g-0}}
%\hfill
% \subfloat[$\gamma=0.42$]{\figureAddLabel{0.315}{time_series/sankey_plot_evi_0.5_f_0.5_g_0.42_h_0_err_0.4_initial__97_01_01_01.pdf}{\bottomText{\scriptsize{Time Steps $t$}}}{}\label{fig:time-series-from-correct-err-0.4-g-0.42}}
%\hfill
%\subfloat[$\gamma=\frac{3}{4}$]{\figureAddLabel{0.45}{time_series/sankey_plot_evi_0.5_f_0.5_g_0.8_h_0_err_0.4_initial__01_01_97_01.pdf}{\bottomText{\scriptsize{Time Steps $t$}}}{}\label{fig:sankey-from-half-correct-err-0.4-g-0.8}}
%\end{minipage}
% \begin{minipage}{0.04\textwidth} 
% \centering
% \marginLeft{\scriptsize{Entropy/Precision}}
% \end{minipage}
%\begin{minipage}{1\textwidth} 
%\subfloat[$\gamma=\frac{1}{3},\evidencerate=0.1, \fusionrate=0.9$]{\figureAddLabel{0.45}{time_series/entropy_plot_evi_0.1_f_0.9_g_0.42_h_0_err_0.4_initial__01_01_97_01.pdf}{\bottomText{\scriptsize{Time Steps $t$}}}{\marginLeft{\scriptsize{Entropy/Precision}}}\label{fig:entropy-from-half-correct-err-0.4-g-0}}
\hfill
% \subfloat[$\gamma=0.42$]{\figureAddLabel{0.315}{time_series/entropy_plot_evi_0.5_f_0.5_g_0.42_h_0_err_0.4_initial__97_01_01_01.pdf}{\bottomText{\scriptsize{Time Steps $t$}}}{}\label{fig:entropy-from-correct-err-0.4-g-0.42}}
\hfill
%\subfloat[$\gamma=\frac{3}{4}$]{\figureAddLabel{0.45}{time_series/entropy_plot_evi_0.5_f_0.5_g_0.8_h_0_err_0.4_initial__01_01_97_01.pdf}{\bottomText{\scriptsize{Time Steps $t$}}}{}\label{fig:entropy-from-half-correct-err-0.4-g-0.8}}
%\end{minipage}

%\end{minipage}

%\begin{minipage}{0.2\linewidth}
%    \includegraphics[width=1\linewidth]{time_series/color_legend.pdf}
%    \includegraphics[width=1\linewidth]{time_series/color_legend_entropy_plot.pdf}
%\end{minipage}
%\caption{Also include these figures for your reference as I feel they are pretty interesting, not sure if we will include them in the paper; All of them are Case 2 (97 percent half correct). a and c are high fusion low evidence scenarios, the converging behaviour is different from others, imprecision is varying around $2$ and seems will finally settle there. b and d, togehter with \ref{fig:sankey-err-0.4-g-0.8}, \ref{fig:entropy-err-0.4-g-0.8} and \ref{fig:sankey-from-correct-err-0.4-g-0.8}, \ref{fig:entropy-from-correct-err-0.4-g-0.8} are high fusion imprecision scenario for all 3 Cases, it looks high fusion imprecision leads the population to the same diversity(have very similar colour components, entropy and precsion values after 100 iterations) regardless initial condition}
%\end{figure}

\revision{In this section, we have presented simulation results for a difference equation model of social learning incorporating varying levels of fusion imprecision. The results show that certain levels of imprecision can improve the accuracy of social learning, particularly when a majority of the population have initially committed to an incorrect belief. While these results are encouraging, it is important to note that the difference equation model concerns the limiting case as the number of agents in the population tends to infinity. Such behaviour may not always be consistent with that of smaller numbers of agents, nor indeed can these models help us understand the effect of changing the population size. To address these limitations, in the following section we introduce agent-based simulations with a limited number of agents to further evaluate the benefits of fusion imprecision.}

\section{Agent-Based Simulation Results}\label{sec:agent-based-simulation}

\revision{Population size has been shown to significantly influence human social learning. Larger populations provide more opportunities for learning and innovation by increasing the diversity of cultural information and reducing the risk of information loss. For example, Henrich’s theoretical models demonstrate that effective population size supports cumulative cultural evolution by buffering against the degradation of cultural knowledge~\citep{henrich2004demography}. Similarly, empirical studies show that larger groups are better at maintaining and improving cultural tools and techniques over time. For a comprehensive review of the relationship between population size and social learning, see~\citep{derex2020cumulative}. Recent studies have been applying agent-based simulation as a straightforward and effective method for modeling social learning under varying population sizes~\citep{lawry2019epistemic,liu2024imprecise}. Typically in agent-based modeling, as the population size grows sufficiently large, the model should in principle produce dynamics close to that of difference equation model described in \Cref{sec:difference-equation-simulations}.}

\tcbset{enhanced,colback=white,colframe=royalblue}
\begin{figure}[!t]
\centering 
\vspace{-6.5em}
    \begin{minipage}{1\linewidth}
     \centering
     \scriptsize{Color bar for Accuracy}\\
     \includegraphics[width=0.5\linewidth]{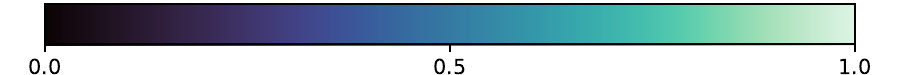}
    \end{minipage} 
    \begin{minipage}{\linewidth}
    \centering
        \begin{minipage}{0.32\linewidth}
            \begin{tcolorbox}[title = \text{\textbf{\scriptsize Case 4} \tiny{ $(0.05,0.05,0.05,0.85)$}}, width = \linewidth]%
                \begin{minipage}{\linewidth}     
                \hspace{-1em}
                    \begin{minipage}{0.02\linewidth}
                        \begin{turn}{90}
                        \centering
                        \tiny $\rho\in(0.02,1]$
                         \end{turn}
                    \end{minipage}
                    \begin{minipage}{0.47\linewidth}{\subfloat{\includegraphics[width=\linewidth]{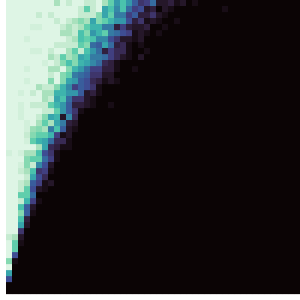}}}       
                    \end{minipage}
                    \begin{minipage}{0.47\linewidth}
                    {\subfloat{\includegraphics[width=\linewidth]{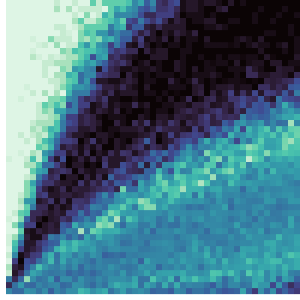}}}
                    \end{minipage}
                    \end{minipage}
                    \begin{minipage}{0.5\linewidth}
                        \centering \tiny $\sigma\in(0.02,1]$
                    \end{minipage}
                    \vspace{-10pt}
                    \begin{minipage}{0.9\linewidth}
                    \vspace{-1.1em}
                    \scriptsize{\textbf{$100$ agents}}
                    \end{minipage} \vspace{-4pt}           
                \tcbline
                 \begin{minipage}{\linewidth}
            \vspace{-10pt}
                       {\subfloat{\includegraphics[width=0.47\linewidth]{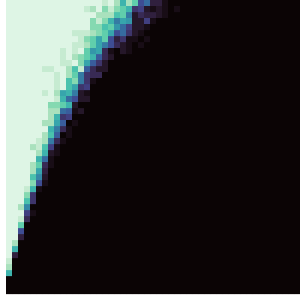}}}
                        {\subfloat{\includegraphics[width=0.47\linewidth]{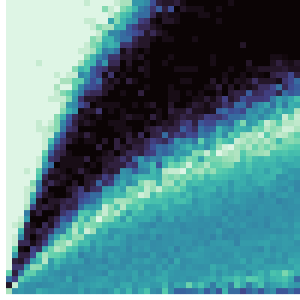}}}
                \end{minipage}
                \scriptsize{\textbf{$200$ agents}}
                \tcbline
                \begin{minipage}{\linewidth}
            \vspace{-10pt}
                {\subfloat{\includegraphics[width=0.47\linewidth]{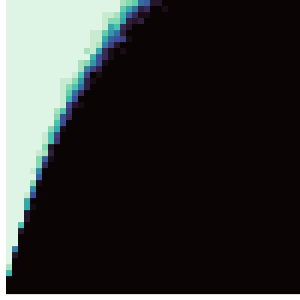}}}
                 {\subfloat{\includegraphics[width=0.47\linewidth]{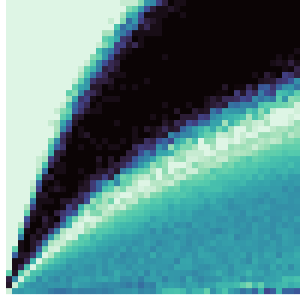}}}
                \end{minipage}
                \scriptsize{\textbf{$400$ agents}}
                \tcbline
             \scriptsize Bias 85\% towards the incorrect belief $\{(0,0)\}$.
            \end{tcolorbox}
        \end{minipage}
\begin{minipage}{0.32\linewidth}

             \begin{tcolorbox}[title = \text{\textbf{\scriptsize Case 5} \tiny$(0.05,0.05,0.85,0.05)$}
             , width = \linewidth]
                \begin{minipage}{\linewidth}
            \vspace{-10pt}
            {\subfloat{\includegraphics[width=0.47\linewidth]{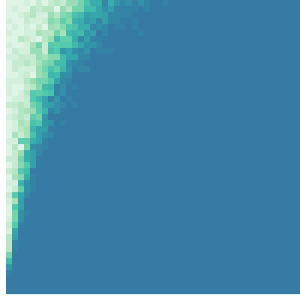}}}
              {\subfloat{\includegraphics[width=0.47\linewidth]{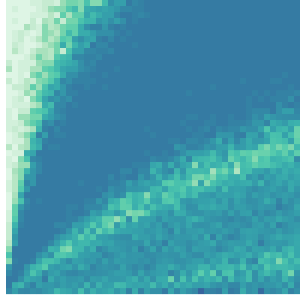}}}
            \end{minipage}
            \scriptsize{\textbf{$100$ agents}}
            \tcbline
            \begin{minipage}{\linewidth}
            \vspace{-10pt}
              {\subfloat{\includegraphics[width=0.47\linewidth]{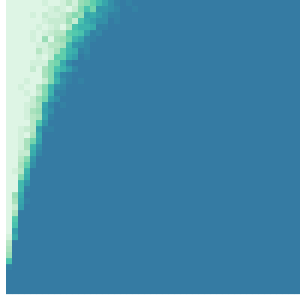}}}
              {\subfloat{\includegraphics[width=0.47\linewidth]{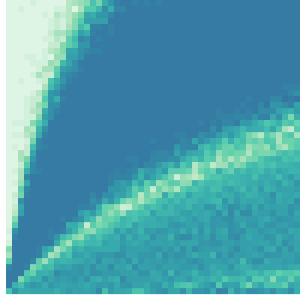}}}
            \end{minipage}
            \scriptsize{\textbf{$200$ agents}}
            \tcbline
            \begin{minipage}{\linewidth}
            \vspace{-10pt}
              {\subfloat{\includegraphics[width=0.47\linewidth]{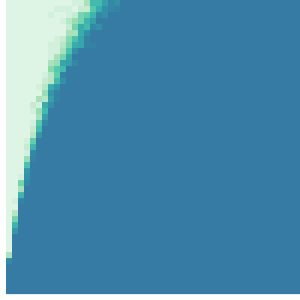}}}
               {\subfloat{\includegraphics[width=0.47\linewidth]{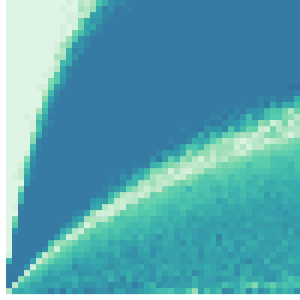}}}
            \end{minipage}
            \scriptsize{\textbf{$400$ agents}}
            \tcbline
             \scriptsize Bias 85\%  $1$-bit incorrect belief $\{(1,0)\}$.
           \end{tcolorbox}
        \end{minipage}
        \begin{minipage}{0.32\linewidth}
            \begin{tcolorbox}[title =  \text{\textbf{\scriptsize Case 6} \tiny$(0.85,0.05,0.05,0.05)$}
            , width = \linewidth]
                \begin{minipage}{\linewidth}
               \vspace{-10pt}                 {\subfloat{\includegraphics[width=0.47\linewidth]{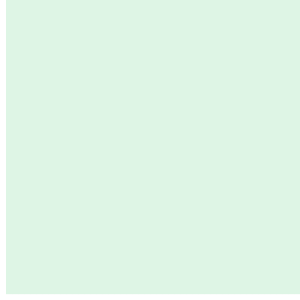}}}
                  {\subfloat{\includegraphics[width=0.47\linewidth]{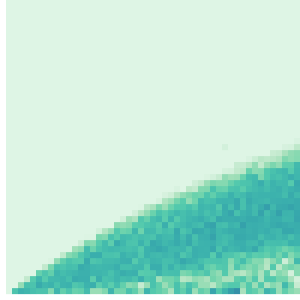}}}
                \end{minipage}
                \scriptsize{\textbf{$100$ agents}}
                \tcblower
                \begin{minipage}{\linewidth}
            \vspace{-10pt}
                  {\subfloat{\includegraphics[width=0.47\linewidth]{biased_agent_simulation/Heatmap_c_rho_vs_sigma_max_likelihood_inter_only_2_an_200_err_41_biased_0.85_0.05_0.05_0.05.pdf}}}
                   {\subfloat{\includegraphics[width=0.47\linewidth]{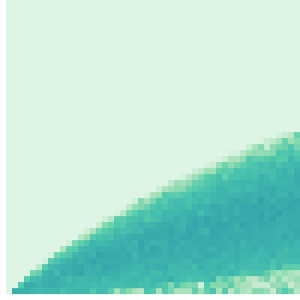}}}
                \end{minipage}
                \scriptsize{\textbf{$200$ agents}}
                \tcbline
                 \begin{minipage}{\linewidth}
            \vspace{-10pt}
                  {\subfloat{\includegraphics[width=0.47\linewidth]{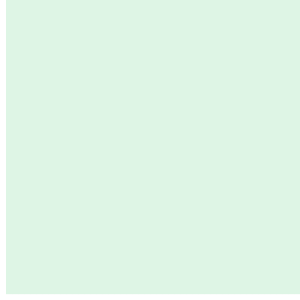}}}
                   {\subfloat{\includegraphics[width=0.47\linewidth]{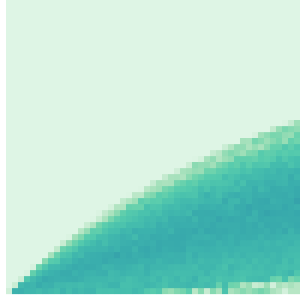}}}
                \end{minipage}
                \scriptsize{\textbf{$400$ agents}}
                \tcbline
                \scriptsize Bias 85\% towards the correct belief $\{(1,1)\}$.
            \end{tcolorbox}
        \end{minipage}
    \end{minipage}
    % \vspace{-3em}
    \begin{minipage}{\linewidth}
    % \vspace{10pt}
    \centering
        \begin{minipage}{0.32\linewidth}
            \begin{tcolorbox}[title =  \text{\textbf{\scriptsize Case 7} \tiny$(0.1,0.1,0.1,0.7)$}
            , width = \linewidth]
                \begin{minipage}{\linewidth}     
                \hspace{-10pt}
                    \begin{minipage}{0.03\linewidth}
                        \begin{turn}{90}
                        \centering
                        \tiny $\rho\in(0.02,1]$
                         \end{turn}
                    \end{minipage}
                    \begin{minipage}{0.47\linewidth}
                    {\subfloat{\includegraphics[width=\linewidth]{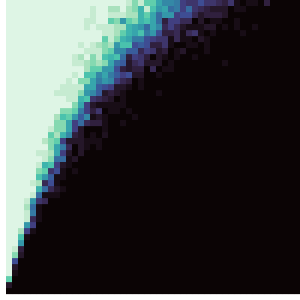}}}       
                    \end{minipage}
                    \begin{minipage}{0.47\linewidth}
                    {\subfloat{\includegraphics[width=\linewidth]{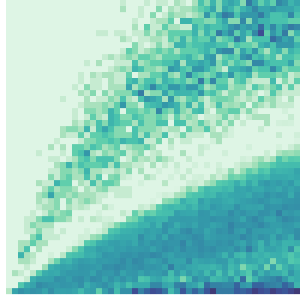}}}
                    \end{minipage}
                    \end{minipage}
                    \begin{minipage}{0.5\linewidth}
                        \centering \tiny $\sigma\in(0.02,1]$
                    \end{minipage}
                    \vspace{-10pt}
                    \begin{minipage}{0.9\linewidth}
                    \vspace{-1.1em}
                        \scriptsize{\textbf{$100$ agents}}
                    \end{minipage}    \vspace{-4pt}          
                \tcbline
                 \begin{minipage}{\linewidth}
            \vspace{-10pt}
                       {\subfloat{\includegraphics[width=0.47\linewidth]{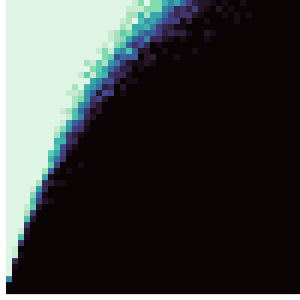}}}
                        {\subfloat{\includegraphics[width=0.47\linewidth]{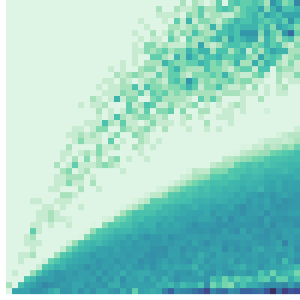}}}
                \end{minipage}
                \scriptsize{\textbf{$200$ agents}}
                \tcbline
                \begin{minipage}{\linewidth}
            \vspace{-10pt}
                {\subfloat{\includegraphics[width=0.47\linewidth]{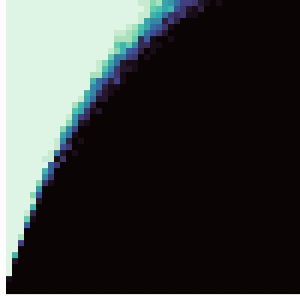}}}
                 {\subfloat{\includegraphics[width=0.47\linewidth]{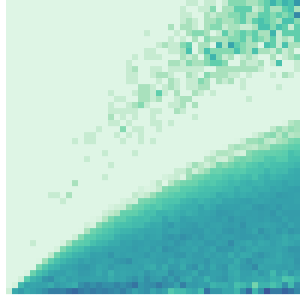}}}
                \end{minipage}
                \scriptsize{\textbf{$400$ agents}}
                \tcbline
             \scriptsize Bias 70\% towards the incorrect belief $\{(0,0)\}$.
            \end{tcolorbox}
        \end{minipage}
        \begin{minipage}{0.32\linewidth}
             \begin{tcolorbox}[title = \text{\textbf{\scriptsize Case 8} \tiny$(0.1,0.1,0.7,0.1)$}
             , width = \linewidth]
                \begin{minipage}{\linewidth}
            \vspace{-10pt}
             {\subfloat{\includegraphics[width=0.47\linewidth]{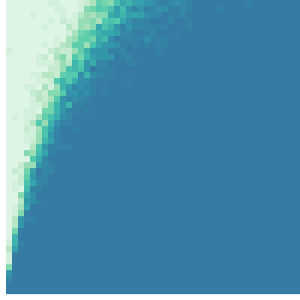}}}
              {\subfloat{\includegraphics[width=0.47\linewidth]{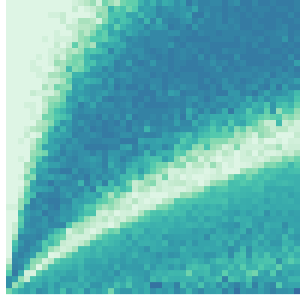}}}
            \end{minipage}
            \scriptsize{\textbf{$100$ agents}}
            \tcbline
            \begin{minipage}{\linewidth}
            \vspace{-10pt}
              {\subfloat{\includegraphics[width=0.47\linewidth]{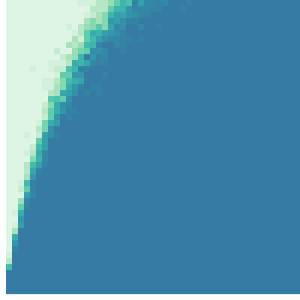}}}
              {\subfloat{\includegraphics[width=0.47\linewidth]{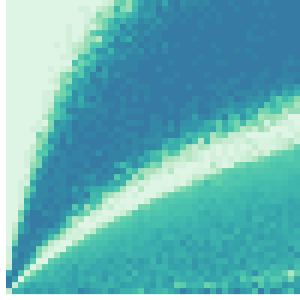}}}
            \end{minipage}
            \scriptsize{\textbf{$200$ agents}}
            \tcbline
            \begin{minipage}{\linewidth}
            \vspace{-10pt}
              {\subfloat{\includegraphics[width=0.47\linewidth]{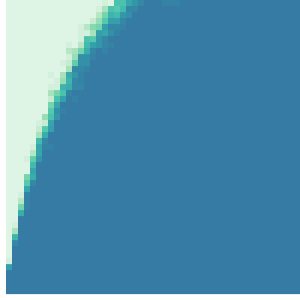}}}
               {\subfloat{\includegraphics[width=0.47\linewidth]{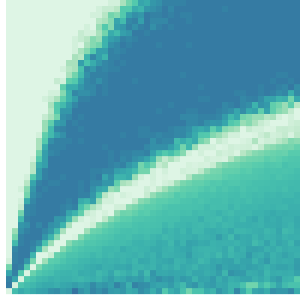}}}
            \end{minipage}
            \scriptsize{\textbf{$400$ agents}}
            \tcbline
            \scriptsize{Bias 70\% towards {$1$-bit} incorrect belief $\{(1,0)\}$}.
           \end{tcolorbox}
        \end{minipage}
        \begin{minipage}{0.32\linewidth}
            \begin{tcolorbox}[title =  \text{\textbf{\scriptsize Case 9} \tiny$(0.7,0.1,0.1,0.1)$}
            , width = \linewidth]
                \begin{minipage}{\linewidth}
                \vspace{-10pt}
                    {\subfloat{\includegraphics[width=0.47\linewidth]{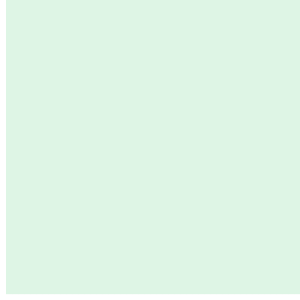}}}
                  {\subfloat{\includegraphics[width=0.47\linewidth]{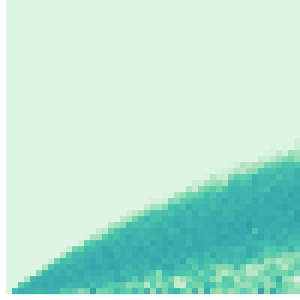}}}
                \end{minipage}
                \scriptsize{\textbf{$100$ agents}}
                \tcblower
                \begin{minipage}{\linewidth}
            \vspace{-10pt}
                  {\subfloat{\includegraphics[width=0.47\linewidth]{biased_agent_simulation/Heatmap_c_rho_vs_sigma_max_likelihood_inter_only_2_an_100_err_41_biased_0.7_0.1_0.1_0.1.pdf}}}
                   {\subfloat{\includegraphics[width=0.47\linewidth]{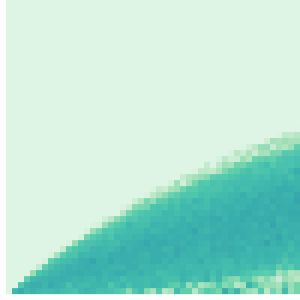}}}
                \end{minipage}
                \scriptsize{\textbf{$200$ agents}}
                \tcbline
                 \begin{minipage}{\linewidth}
            \vspace{-10pt}
                  {\subfloat{\includegraphics[width=0.47\linewidth]{biased_agent_simulation/Heatmap_c_rho_vs_sigma_max_likelihood_inter_only_2_an_100_err_41_biased_0.7_0.1_0.1_0.1.pdf}}}
                   {\subfloat{\includegraphics[width=0.47\linewidth]{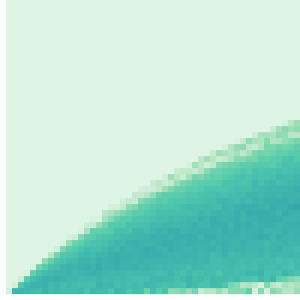}}}
                \end{minipage}
                \scriptsize{\textbf{$400$ agents}}
                \tcbline
                \scriptsize Bias 70\% towards the correct belief $\{(1,1)\}$.
            \end{tcolorbox}
        \end{minipage}
    \end{minipage}
    \vspace{-10pt}
    \caption{Heatmaps showing average population accuracy for $(\fusionrate,\evidencerate)$ parameter space at steady state or after $20,000$ iterations for population size $n\in\{100,200,400\}$ and $\errorrate=0.4$. For each pair: \textbf{Left:} $\fthreshold=0$; \textbf{Right:} $\fthreshold=\frac{1}{3}$. The sample axis labels are provided on the top left figure. \revision{Imprecise fusion $\gamma =\frac{1}{3}$ can enhance overall performance in Cases 4, 5, 7, and 8, particularly when the evidence rate is high. The benefits become more significant as the population size increases and bias decreases.}}
    \label{fig:agent-based-err-0.4}
\end{figure}

In this section, we present the results from agent-based simulations for the proposed imprecision fusion model with fusion thresholds $\fthreshold = 0$ and $\fthreshold = \frac{1}{3}$ (where $\frac{1}{3} \in \left[\frac{1}{3}, \frac{1}{2}\right)$). The simulations help to validate the results of the difference equations experiments in the case when the population size is large and provides insight into the effect that varying the population size has on learning performance. Specifically, we will consider agent-based simulations with population sizes $k\in\{100,200,400\}$, language size $n=2$ propositions, and fusion and evidence rates $(\fusionrate, \evidencerate) \in (0,1)^2$, and error rates $\errorrate\in\{0.2,0.3,0.4\}$. Furthermore, we consider six different scenarios which are variations on those introduced in \Cref{sec:difference-equation-simulations}. Initially, we introduce the following three cases. \textbf{Case 4:} initial beliefs are selected at random such that an agent adopts the incorrect belief, i.e $B=\{s_4\}$, with probability $0.85$, and adopts each of the other precise beliefs with probability $0.05$; \textbf{Case 5:} initial beliefs are selected at random such that an agent adopts one of the half correct beliefs, i.e. $B=\{s_2\}$ or $B=\{s_3\}$, with probability $0.85$, and adopts each of the other precise beliefs with probability $0.05$; \textbf{Case 6:} initial beliefs are selected at random such that an agent adopts the correct belief, i.e. $B=\{ s_4\}$, with probability $0.85$, and adopts each of the other precise beliefs with probability $0.05$. The parameter settings for these cases are detailed in \Cref{table:parameters} for better clarity. We apply a lower initial bias in these scenarios compared to those considered in \Cref{sec:difference-equation-simulations}, because highly biased probabilities for the allocation of initial beliefs can, with non-zero probability, result in very homogeneous populations when the population size is bounded, and which in turn severely limits both evidential learning and fusion for all agents. For example, if we initially assign probabilities of $(0.01,\ 0.01,\ 0.01,\ 0.97)$ to the four beliefs ${B_1,\ B_2,\ B_3,\ B_4}$, then in a population of $100$ agents there is a $4.7\%$ probability that only $B_4$ is present, a $25.5\%$ probability that two of ${B_1,\ B_2,\ B_3}$ are missing, and a $44.9\%$ probability that one is missing. 

\Cref{fig:agent-based-err-0.4} shows agent-based simulation results for varying population sizes and initial conditions in the form of accuracy heat maps for $(\sigma,\rho)$ parameter space. In order to take account of the inherent stochasticity of the model, each parameter combination is simulated 50 times and the accuracy results averaged over those indpendent runs. Consistent with the difference equation simulation results in \Cref{fig:diff-eq-sim}, we see that when the fusion rate is low relative to the evidence rate the learning accuracy is robust across different initial conditions and population sizes. In particular,  increasing the population size does not significantly improve the learning accuracy in such cases. In other regions of the $(\fusionrate,\evidencerate)$ parameter space, when there is precise fusion, i.e. $\fthreshold=0$, the population tends to reach a consensus aligned with initial biases, and which  matches the stability regions ($R_2$ and $R_4$) shown in \Cref{fig:stability-regions-err-0.4-g-0.42}.  In addition, the agent-based simulations show that imprecise fusion outperforms precise fusion when the population belief is biased toward an incorrect belief. However, we note that this difference in performance is greater for larger populations that better approximate the behaviour predicted by the difference equation models.

We now consider three additional scenarios in which we further reduce the initial bias in the population; \textbf{Case 7:}  initial beliefs are selected at random such that an agent adopts the incorrect belief, i.e $B=\{s_4\}$, with probability $0.7$, and adopts each of the other precise beliefs with probability $0.1$;  \textbf{Case 8:} initial beliefs are selected at random such that an agent adopts one of the half correct beliefs, i.e. $B=\{s_2\}$ or $B=\{s_3\}$, with probability $0.7$, and adopts each of the other precise beliefs with probability $0.1$; \textbf{Case 9:} initial beliefs are selected at random such that an agent adopts the correct belief, i.e. $B=\{ s_4\}$, with probability $0.7$, and adopts each of the other precise beliefs with probability $0.1$.

\begin{figure}[t]
    \begin{minipage}{1\linewidth}
     \centering
     \scriptsize{Color bar for Accuracy}\\
     \includegraphics[width=0.7\linewidth]{diff_eq_sim/accuracy_colorbar_mako.pdf}
    \end{minipage}  
    \centering
    \subfloat[$\errorrate=0.2$, Case 4]{{\includegraphics[width=0.23\linewidth]{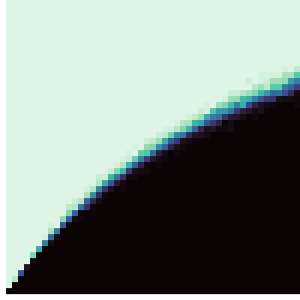}}
    {\includegraphics[width=0.23\linewidth]{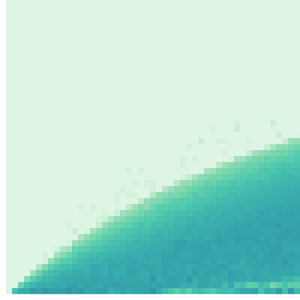}}}
    \hfill
    \subfloat[$\errorrate=0.2$, Case 5]{\includegraphics[width=0.23\linewidth]{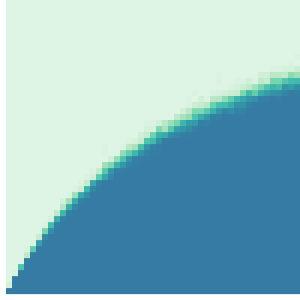}
    {\includegraphics[width=0.23\linewidth]{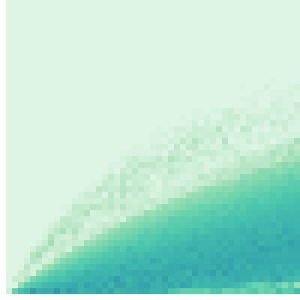}}}

    \subfloat[$\errorrate=0.3$, Case 4]{{\includegraphics[width=0.23\linewidth]{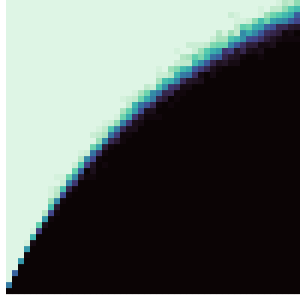}}
    {\includegraphics[width=0.23\linewidth]{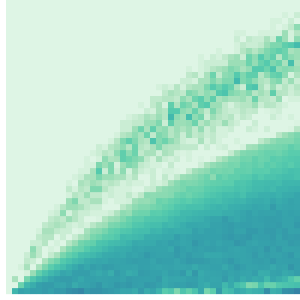}}}
    \hfill
     \subfloat[$\errorrate=0.3$, Case 5]{{\includegraphics[width=0.23\linewidth]{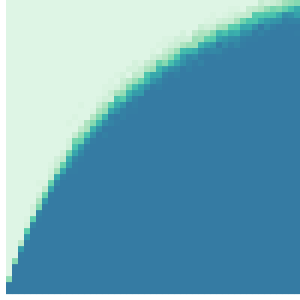}}
    {\includegraphics[width=0.23\linewidth]{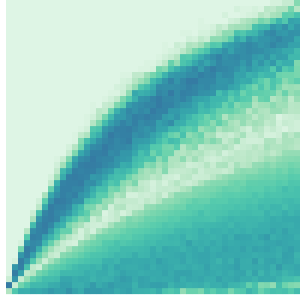}}}

    \caption{Heatmaps of average accuracy for population size $k=400$, and different error rates. For each pair: \textbf{Left:} precise fusion $\fthreshold=0$; \textbf{Right:} imprecise fusion $\fthreshold=\frac{1}{3}$. The sample axis labels are provided on the top left figure in \Cref{fig:agent-based-err-0.4} .}
    \label{fig:agent-based-err}
\end{figure}

The agent-based simulation results for \textbf{Case 7}, \textbf{Case 8} and \textbf{Case 9}  are shown in  \Cref{fig:agent-based-err-0.4}. Notice that adopting these less biased initial conditions does not result in significant improvements in performance when precise fusion is used. However, the benefits of imprecise fusion are now more clearer. As for \textbf{Case 4}, \textbf{Case 5} and \textbf{Case 6}, we observe that the learning performance improves as the population size increases, particularly when imprecise fusion is used. Larger numbers of agents increase the overall resilience of the population, allowing it to better recover from biased initial conditions. These results highlight the importance of population size in maintaining robustness, as smaller populations may struggle to correct initial biases effectively. Moreover, imprecise fusion becomes increasingly advantageous in larger populations, where it significantly outperforms precise fusion in scenarios where the population’s initial belief is skewed towards incorrect information. These findings suggest that the combination of a larger population and imprecise fusion can help mitigate the detrimental effects of bias, thus improving overall system stability and learning accuracy. Furthermore, we consider two scenarios with lower error rates, $\errorrate\in\{0.2,0.3\}$
, in \Cref{fig:agent-based-err}). As anticipated from the difference equation model (see \Cref{fig:benefits-against-error}), the absolute benefits of imprecise fusion become increasingly significant when the error rate falls within the moderate range of 
 to . This trend substantiates that imprecise fusion significantly enhances the system’s resilience and adaptability under moderate error conditions.

\section{Discussion and Conclusion}\label{sec:conclusion}

We have investigated social learning for a propositional model of belief in which an agent's belief is the set of states that it believes to possibly be the true state. Agents learn from two distinct sources; directly from the environment using belief updating, and by applying a fusion operator to combine their beliefs with those of others in the population. Furthermore, we allow for the fusion operator to have varying levels of imprecision, where more imprecise operators tend to result in more imprecise beliefs as measured by the cardinality of belief sets. Our findings show that when fusion rates are low compared to evidence rates, the system has good learning accuracy across various levels of imprecision and initial conditions. These regions of the $(\sigma,\rho)$ parameter space tend to be those in which the correct belief is the only stable fixed point. In contrast, when fusion is precise and when the fusion rate is high, the system converges to states aligned with initial biases. However, in such scenarios imprecise fusion can improve learning accuracy by enabling the population to overcome an initial bias and reach consensus on the true state of the world, particularly if the population size is relatively large.

The models discussed in this paper help to highlight the potential benefits of imprecision in social learning when beliefs are expressed in a richer propositional language rather than as a single numerical or truth value. In this context, imprecision in the fusion operator ensures that disagreement between individuals results in them adopting more imprecise beliefs, and which in turns allows them to be more open minded whenever new evidence becomes available. This  provides insights into the effectiveness of social learning in highly connected human societies, where collective decision-making often involves multiple factors, and where beliefs and opinions tend to be expressed in a propositional, rather than a numerical, form. In addition, these models may also help to inform the design and analysis of Multi-Agent Systems (MAS) undertaking collective decision-making tasks in which options are characterised by multiple discrete states of the world.

There are a number of important limitations to the models we have introduced and which provide interesting potential avenues for future research in this area. For instance, both the difference equation models and the agent-based simulations assume a well-mixed population, and consider only the case in which the agents all belong to a fully connected social network so that any two agents are free to interact at any given time. However, in practice social learning is often much more constrained. For instance, human social interactions, while increasingly widespread, are nonetheless restricted to limited groups of friends, acquaintances or other sources, even in an online setting. In addition, robots and autonomous systems operating in complex physical environments are typically constrained in their interactions at any time by their relative positions and communication limitations. There is a natural tendency in all of these examples to assume that restricting interactions between agents will have a negative effect on social learning. However, there is a growing body of research that suggests that fully connected social networks are not always optimal and that limiting direct interactions to a smaller set of network neighbours can sometimes result in more accurate and robust learning performance. Future research could benefit from exploring the proposed imprecise fusion-based propositional model for other types of social network such as, for example, small world networks.

In its general form the propositional framework we have described can be used to model social learning with any number of propositions. However, in this paper we have only investigated the two propositional case which is the simplest non-trivial case. In future work it would therefore be interesting to extend our analysis to languages with $n>2$, to see, for example, if the benefits of imprecise fusions can be extended in an analogous way to those cases. However, there are significant challenges in carrying out this analysis, at least for large values of $n$. In particular, the dimensionality of the associated difference equation models is exponential in $2^n$. In other words, $n$ propositions have $2^n$ different allocations of truth values and hence there are $2^{2^n}-1$ possible beliefs that agents can adopt. Clearly, then, there are inherent limits to the size of the propositional language for which we can analyse social learning using the types of models described in this paper, but it may nonetheless be insightful to carry out a detailed analysis of slightly larger languages with $n=3$ or $n=4$ propositions.
 

\appendix
% \section{Example Appendix Section}
% \appendices
\section{Updating Matrix}\label{ap:transition-matrices}
\begin{gather}\label{eq:updating-matrix} 
\scriptsize
\begin{sideways}
$\begin{bmatrix}
\begin{smallmatrix}
1 & 0 & 0 & 0 & \rho \left(1-\epsilon\right)^2 & \rho \left(1-\epsilon\right)^2 & \rho \left(1-\epsilon\right)^2 & 0 & 0 & 0 & \rho \left(1-\epsilon\right)^2 & \rho \left(1-\epsilon\right)^2 & \rho \left(1-\epsilon\right)^2 & 0 & \rho \left(1-\epsilon\right)^2 \\
0 & 1 & 0 & 0 & \epsilon \rho \left(1 - \epsilon\right) & 0 & 0 & \epsilon \rho \left(1 - \epsilon\right) & \epsilon \rho \left(1 - \epsilon\right) & 0 & \epsilon \rho \left(1 - \epsilon\right) & \epsilon \rho \left(1 - \epsilon\right) & 0 & \epsilon \rho \left(1 - \epsilon\right) & \epsilon \rho \left(1 - \epsilon\right) \\
0 & 0 & 1 & 0 & 0 & \epsilon \rho \left(1 - \epsilon\right) & 0 & \epsilon \rho \left(1 - \epsilon\right) & 0 & \epsilon \rho \left(1 - \epsilon\right) & \epsilon \rho \left(1 - \epsilon\right) & 0 & \epsilon \rho \left(1 - \epsilon\right) & \epsilon \rho \left(1 - \epsilon\right) & \epsilon \rho \left(1 - \epsilon\right) \\
0 & 0 & 0 & 1 & 0 & 0 & \epsilon^2 \rho & 0 & \epsilon^2 \rho & \epsilon^2 \rho & 0 & \epsilon^2 \rho & \epsilon^2 \rho & \epsilon^2 \rho & \epsilon^2 \rho \\
0 & 0 & 0 & 0 & \epsilon \rho - \rho + 1 & 0 & 0 & 0 & 0 & 0 & 0 & 0 & 0 & 0 & 0 \\
0 & 0 & 0 & 0 & 0 & \epsilon \rho - \rho + 1 & 0 & 0 & 0 & 0 & 0 & 0 & 0 & 0 & 0 \\
0 & 0 & 0 & 0 & 0 & 0 & 2 \epsilon \rho \left(1 - \epsilon\right) - \rho + 1 & 0 & 0 & 0 & 0 & 0 & 0 & 0 & 0 \\
0 & 0 & 0 & 0 & 0 & 0 & 0 & \rho \left(\epsilon^2 + \left(1-\epsilon\right)^2\right) - \rho + 1 & 0 & 0 & 0 & 0 & 0 & 0 & 0 \\
0 & 0 & 0 & 0 & 0 & 0 & 0 & 0 & \rho \left(1 - \epsilon\right) - \rho + 1 & 0 & 0 & 0 & 0 & 0 & 0 \\
0 & 0 & 0 & 0 & 0 & 0 & 0 & 0 & 0 & \rho \left(1 - \epsilon\right) - \rho + 1 & 0 & 0 & 0 & 0 & 0 \\
0 & 0 & 0 & 0 & 0 & 0 & 0 & 0 & 0 & 0 & \epsilon^2 \rho - \rho + 1 & 0 & 0 & 0 & 0 \\
0 & 0 & 0 & 0 & 0 & 0 & 0 & 0 & 0 & 0 & 0 & \epsilon \rho \left(1 - \epsilon\right) - \rho + 1 & 0 & 0 & 0 \\
0 & 0 & 0 & 0 & 0 & 0 & 0 & 0 & 0 & 0 & 0 & 0 & \epsilon \rho \left(1 - \epsilon\right) - \rho + 1 & 0 & 0 \\
0 & 0 & 0 & 0 & 0 & 0 & 0 & 0 & 0 & 0 & 0 & 0 & 0 & \rho \left(1-\epsilon\right)^2 - \rho + 1 & 0 \\
0 & 0 & 0 & 0 & 0 & 0 & 0 & 0 & 0 & 0 & 0 & 0 & 0 & 0 & 1 - \rho
\end{smallmatrix}
\end{bmatrix}
$ 
\end{sideways}
\end{gather}
\section{Fusion Matrix}
To concisely express the fusion transition matrix, we represent it as the sum of a diagonal matrix $diag(D)$, which represents the probabilities of beliefs remaining unchanged by fusion, and a partial fusion matrix $F^\prime$ 
that contains the remaining probabilities.The fusion matrix is $F=F^\prime+diag(D)$, where $D$ and $F^\prime$ are as follows: in $D$ and $F^\prime$, $P_{i+\dots+j}$ represents $P_i+\dots+P_j$, where summing the indices is an abuse of notation intended to be a concise way of expressing the sum of the associated probabilities: 
% \vspace{-1em}
\begin{gather}\label{eq:fusion-matrix-diag} 
% 对角线向量
D = \begin{bmatrix}
    \fusionrate P_{1+5+11+15+12+6+13+7} +1 - \fusionrate  \\
    \fusionrate P_{5+11+15+12+2+8+14+9} +1 - \fusionrate \\
    \fusionrate P_{11+15+6+13+8+14+3+10} +1 - \fusionrate \\
    \fusionrate P_{15+12+13+7+14+9+10+4} +1 - \fusionrate \\
    \fusionrate P_{5+11+15+12} +1 - \fusionrate \\
    \fusionrate P_{11+15+6+13} +1 - \fusionrate \\
    \fusionrate P_{15+12+13+7} +1 - \fusionrate \\
    \fusionrate P_{11+15+8+14} +1 - \fusionrate \\
    \fusionrate P_{15+12+14+9} +1 - \fusionrate \\
    \fusionrate P_{15+13+14+10} +1 - \fusionrate \\
    \fusionrate P_{11+15} +1 - \fusionrate \\
    \fusionrate P_{15+12} +1 - \fusionrate \\
    \fusionrate P_{15+13} +1 - \fusionrate \\
    \fusionrate P_{15+14} +1 - \fusionrate \\
    \fusionrate P_{15} +1 - \fusionrate
\end{bmatrix}
\end{gather}

% 替换对角线为 0 的矩阵
\begin{gather}\label{eq:fusion-matrix-other} 
\begin{sideways}
$F^\prime=\fusionrate\cdot$
$\begin{bmatrix}
\begin{smallmatrix}
    0 & 0 & 0 & 0 &  P_{1+6+13+7} &  P_{1+5+12+7} &  P_{1+5+11+6} & 0 & 0 & 0 &  P_{1+7} &  P_{1+6} &  P_{1+5} & 0 &  P_{1}\\
    0 & 0 & 0 & 0 &  P_{2+8+14+9} & 0 & 0 &  P_{5+12+2+9} &  P_{5+11+2+8} & 0 &  P_{2+9} &  P_{2+8} & 0 &  P_{5+2} &  P_{2}\\
    0 & 0 & 0 & 0 & 0 &  P_{8+14+3+10} & 0 &  P_{6+13+3+10} & 0 &  P_{11+6+8+3} &  P_{3+10} & 0 &  P_{8+3} &  P_{6+3} &  P_{3}\\
    0 & 0 & 0 & 0 & 0 & 0 &  P_{14+9+10+4} & 0 &  P_{13+7+10+4} &  P_{12+7+9+4} & 0 &  P_{10+4} &  P_{9+4} &  P_{7+4} &  P_{4}\\
     P_{2} &  P_{1} & 0 & 0 & 0 & 0 & 0 & 0 & 0 & 0 &  P_{5+12} &  P_{5+11} & 0 & 0 &  P_{5}\\
     P_{3} & 0 &  P_{1} & 0 & 0 & 0 & 0 & 0 & 0 & 0 &  P_{6+13} & 0 &  P_{11+6} & 0 &  P_{6}\\
     P_{4} & 0 & 0 &  P_{1} & 0 & 0 & 0 & 0 & 0 & 0 &  P_{13+7} &  P_{12+7} & 0 &  P_{7}\\
    0 &  P_{3} &  P_{2} & 0 & 0 & 0 & 0 & 0 & 0 & 0 &  P_{8+14} & 0 & 0 &  P_{11+8} &  P_{8}\\
    0 &  P_{4} & 0 &  P_{2} & 0 & 0 & 0 & 0 & 0 & 0 &  P_{14+9} & 0 &  P_{12+9} &  P_{9}\\
    0 & 0 &  P_{4} &  P_{3} & 0 & 0 & 0 & 0 & 0 & 0 & 0 & 0 &  P_{14+10} &  P_{13+10} &  P_{10}\\
     P_{8} &  P_{6} &  P_{5} & 0 &  P_{3} &  P_{2} & 0 &  P_{1} & 0 & 0 & 0 & 0 & 0 & 0 &  P_{11}\\
     P_{9} &  P_{7} & 0 &  P_{5} &  P_{4} & 0 &  P_{2} & 0 &  P_{1} & 0 & 0 & 0 & 0 & 0 &  P_{12}\\
     P_{10} & 0 &  P_{7} &  P_{6} & 0 &  P_{4} &  P_{3} & 0 & 0 &  P_{1} & 0 & 0 & 0 & 0 &  P_{13}\\
    0 &  P_{10} &  P_{9} &  P_{8} & 0 & 0 & 0 &  P_{4} &  P_{3} &  P_{2} & 0 & 0 & 0 & 0 &  P_{14}\\
     P_{14} &  P_{13} &  P_{12} &  P_{11} &  P_{10} &  P_{9} &  P_{8} &  P_{7} &  P_{6} &  P_{5} &  P_{4} &  P_{3} &  P_{2} &  P_{1} & 0\\
\end{smallmatrix}
\end{bmatrix}
$ 
\end{sideways}
\end{gather}

% \section{}

\section*{Acknowledgment}
The agent-based model applied in this paper was developed as part of Zixuan Liu's PhD research, which was supported by a joint studentship from the University of Bristol and the China Scholarship Council (Grant No. 201908150152). Michael Crosscombe is supported by JSPS KAKENHI JP23KF0108 Grant-in-Aid for Research Fellows.

%%       \cite{<label>} ==> [1]

\bibliographystyle{elsarticle-num} 
\bibliography{reference}




% %% For numbered reference style

%   \textit{\LaTeX: a document preparation system},
%   Addison Wesley, Massachusetts,

% \end{thebibliography}
\end{document}